\documentclass[reprint,superscriptaddress,amsmath,amssymb,pra,floatfix]{revtex4-2}
\usepackage{graphicx}% Include figure files
\usepackage{bm}% bold math
\usepackage{amsmath}
\usepackage{xcolor}
\usepackage[colorlinks,linkcolor=blue,citecolor=blue,urlcolor=blue]{hyperref}% add hypertext capabilities

\begin{document}

\title{Quantized pumping in disordered nonlinear Thouless pumps}% Force line breaks with \\
\author{Abhijit P. Chaudhari}
\email{apc6283@psu.edu}
\affiliation{Department of Physics, The Pennsylvania State University, University Park, Pennsylvania 16802, USA}
\author{Marius J\"urgensen}
\affiliation{Department of Physics, The Pennsylvania State University, University Park, Pennsylvania 16802, USA}
\affiliation{Department of Physics, Stanford University, Stanford, California, 94305, USA}
\author{Mikael C. Rechtsman}
\email{mcrworld@gmail.com}
\affiliation{Department of Physics, The Pennsylvania State University, University Park, Pennsylvania 16802, USA}

\makeatletter
\def\maketitle{\@author@finish
\title@column\titleblock@produce
\suppressfloats[t]}
\makeatother

\date{\today}
\begin{abstract}
We investigate the dynamics of nonlinear optical Thouless pumps in the presence of disorder, using optical waveguide arrays.  It was previously known that the displacement of solitons in Thouless pumps is quantized and may exhibit integer and fractional transport over the course of the pump cycle. Here, we demonstrate that, in disordered nonlinear pumps, quantization may be maintained despite the presence of disorder, even though it would not be in the linear domain. Moreover, nonlinearity allows pumps to be executed more quickly (i.e., less adiabatically).  This may serve as a design principle for integrated non-reciprocal devices based on temporal modulation.
\end{abstract}

\maketitle

\textit{Introduction}---Prompted by the discovery of the integer quantum Hall effect~\cite{PhysRevLett.45.494}, several theoretical explanations were proposed to understand the quantization of conductance~\cite{PhysRevB.23.5632,avron1983homotopy}. D.J. Thouless presented a model involving a one-dimensional system with adiabatically varying time-periodic potential that pumps a quantized amount of charge per period \cite{thouless1982quantized}. These models are mathematically equivalent to Chern insulators in two dimensions and hence serve as an explanation for the quantized conductance observed in the integer quantum Hall effect~\cite{PhysRevLett.45.494,RevModPhys.82.1959,asboth2016short,citro2023thouless}. Such models are known as ``Thouless pumps" and have become a platform for the study of topological physics in fields beyond condensed matter physics and have been realized in a wide range of experimental platforms such as ultracold atomic systems~\cite{lohse2016thouless,nakajima2016topological,PhysRevLett.116.200402,lohse2018exploring,PhysRevLett.129.053201}, photonics~\cite{PhysRevLett.109.106402,verbin2013observation,PhysRevB.91.064201,ke2016topological,zilberberg2018photonic,fedorova2020observation,sun2022non,wang2022two,PhysRevA.107.033501,sun2024two,doi:10.1073/pnas.2411793121,jürgensen2025quantizeddynamicalpumpingdissipation,tao2025thouless} and mechanical experiments~\cite{grinberg2020robust,PhysRevResearch.6.023010,52yh-mlfm}. It has also been shown that disorder and interactions in fermionic Thouless pumps do not destroy quantized transport as long as they do not close the many-body gap~\cite{niu1985quantized}. However, the argument does not hold for non-interacting (linear) bosonic Thouless pumps (as experimentally studied, e.g. in ~\cite{cerjan2020thouless,nakajima2021competition, liu2025interplay}), because even an infinitesimal amount of disorder leads to mixing of populations in different states within the relevant band, hence destroying the quantized transport. 

Most studies to date on Thouless pumping have been in the linear domain (for a recent review on Thouless pumping, see Ref.~\cite{citro2023thouless}). Recently, theoretical~\cite{PhysRevA.95.063630,PhysRevB.98.245148,PhysRevResearch.5.013020,huang2024topological,hu2024pumping,4d5s-n4gn,PhysRevA.111.033306,ye2025thouless,PhysRevLett.134.093801} and some experimental works~\cite{PhysRevLett.117.213603,walter2023quantization,PhysRevX.14.021049,kiefer2025protectedquantumgatesusing} have explored the interplay between pumping and interparticle interactions. In the domain of photonics, photon-photon interactions take the form of optical nonlinearity in the mean-field (many photon) limit, and therefore structured nonlinear optical systems are natural platforms in which to study this interplay. In a series of recent papers using optical waveguide arrays, it was found that spatial solitons in Thouless pumps exhibit quantized displacement in a similar spirit to Wannier functions~\cite{jurgensen2021quantized,jurgensen2023quantized,fu2022nonlinear}. This is perhaps counterintuitive because the soliton is a very different object than a Wannier function and does not equally populate each state in a band. Although soliton displacement has been proven to be rigorously quantized in the low-power regime~\cite{jurgensen2022chern,mostaan2022quantized}, it was found that fractional quantization can emerge at intermediate powers~\cite{fu2022nonlinear,jurgensen2023quantized}. Topological transitions in soliton displacement occur as a result of nonlinear bifurcations, and such bifurcations have been shown to be equivalent to the closing of a many-body gap \cite{bohm2025quantumtheoryfractionaltopological}. Further work has also theoretically studied nonlinearly-induced Thouless pumping ~\cite{tao2024nonlinearityinducedthoulesspumpingsolitons,96f5-qszj}. A distinctive feature of the Thouless pumping of solitons is that they need not be as adiabatic as in the linear case because of the spectral separation between the soliton energy and the linear band. Taken together, quantization of soliton displacement, as well as the fact of near-quantization even when deviating from adiabatic pumping, suggests that nonlinear Thouless pumps should therefore be intrinsically more robust to disorder than linear ones~\cite{cao2024nonlinear,jürgensen2025quantizeddynamicalpumpingdissipation}.   

Here, we show theoretically and experimentally that nonlinear photonic Thouless pumps are highly robust to disorder, improving upon the linear case both in terms of the degree of disorder and the speed with which the pump may be executed. We use coupled waveguide arrays, governed by the discrete nonlinear Schr\"odinger equation in the presence of a focusing Kerr nonlinearity. We start by numerically studying the simplified case in which a soliton pumps across a single defect in a waveguide array and find that the quantized displacement of the soliton remains intact for defect strengths below a critical value. This conclusion remains true even when the soliton is pumped through a disordered region consisting of multiple single-site defects, implying that the topological nature of the pump protects the soliton displacement. We experimentally demonstrate this effect by exciting solitons in coupled waveguide arrays fabricated using femtosecond laser writing~\cite{davis1996writing,szameit2010discrete} and observe robust soliton pumping in the presence of disorder.

\begin{figure}[ht]
\includegraphics[width = 8.5cm]{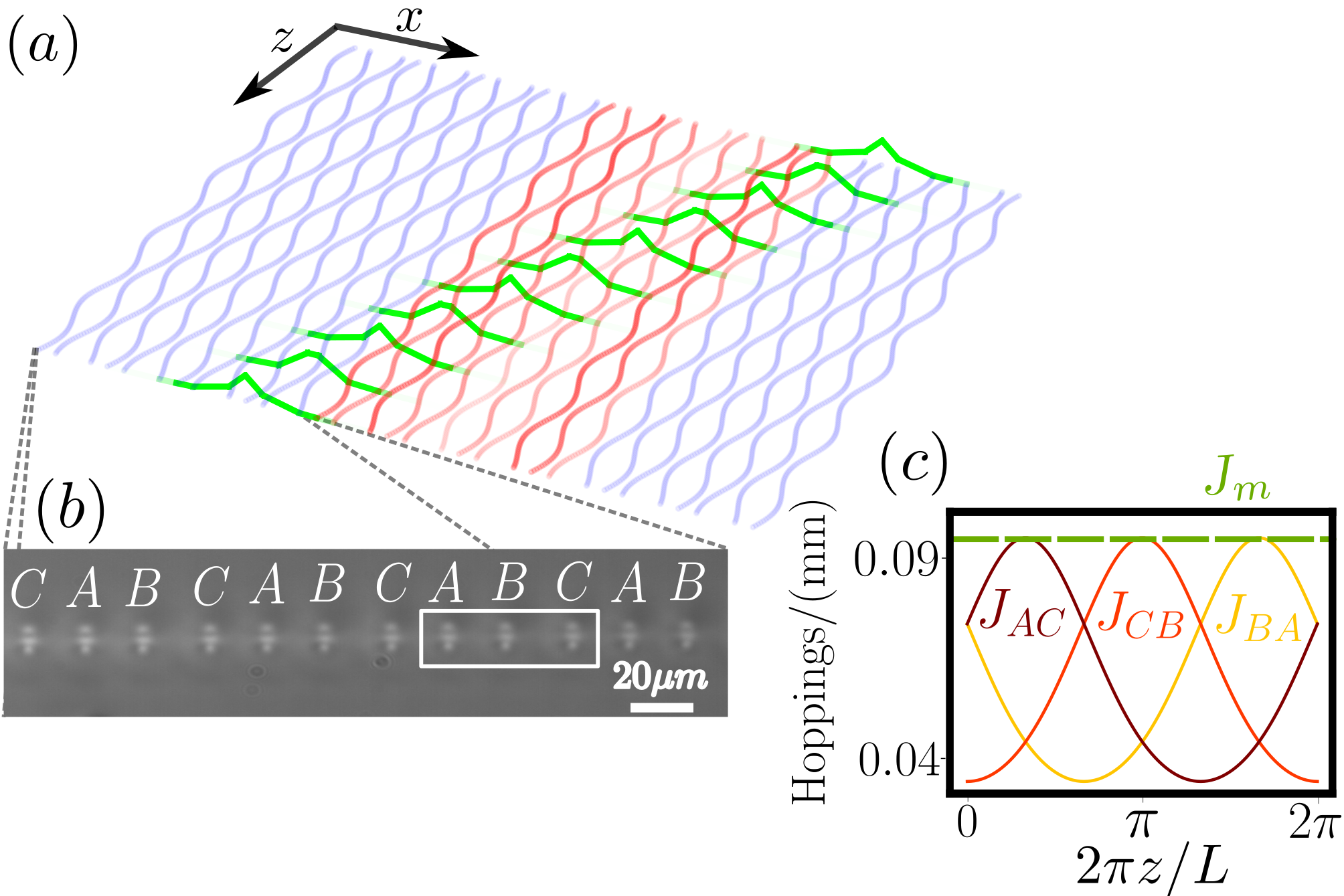}
\caption{\textbf{Nonlinear Thouless pumping in an optical waveguide array.} ($a$) Schematic representation of a disordered Thouless pump implemented in an array of coupled waveguides, with a soliton (green) pumping through the disordered region (red). ($b$) White light image of the output facet of an experimentally fabricated waveguide array. ($c$) Variation of hopping strength between neighboring waveguides as a function of propagation distance $z$ for one period. $J_m$ represents maximum hopping strength between two neighboring waveguides.}
\label{schematics}
\end{figure}

\textit{Numerical simulations.}--- We model coupled single-mode waveguide arrays using the discrete nonlinear Schr\"odinger equation: 
\begin{align}
\label{DNLSE}
i \partial_z \phi_n(z) &= J_n(z) \phi_{n+1}(z) + J_{n-1}(z) \phi_{n-1}(z) \nonumber \\
&\quad - g |\phi_n(z)|^2 \phi_n(z) + V_n \phi_n(z)
\end{align}
where the propagation distance $z$ plays the role of time, $n$ is a waveguide index (site), $\phi_{n}(z)$ is the wavefunction (electric field amplitude) at site $n$, $g$ is the Kerr nonlinearity strength, $V_n$ is the potential (capturing defects) and $J_n(z)$ is the hopping strength between neighboring sites. In the following, we denote the maximum hopping strength between the neighboring waveguides by $J_m$ and $g\sum_n|\phi_n(z)|^2/J_m = gP/J_m$ as the power. To implement Thouless pumping in coupled waveguide arrays (see Fig.~\ref{schematics} ($a$)), the $n^{\mathrm{th}}$ waveguide is modulated adiabatically along $z$ according to:
\begin{equation}
\label{waveguide_pumping}
x_{n}(z) = nd + \delta d\sin\left(\frac{2\pi n}{3} + \frac{2\pi}{L} z - \pi\right).
\end{equation}

\begin{figure}[ht]
\includegraphics[width = 8.5cm]{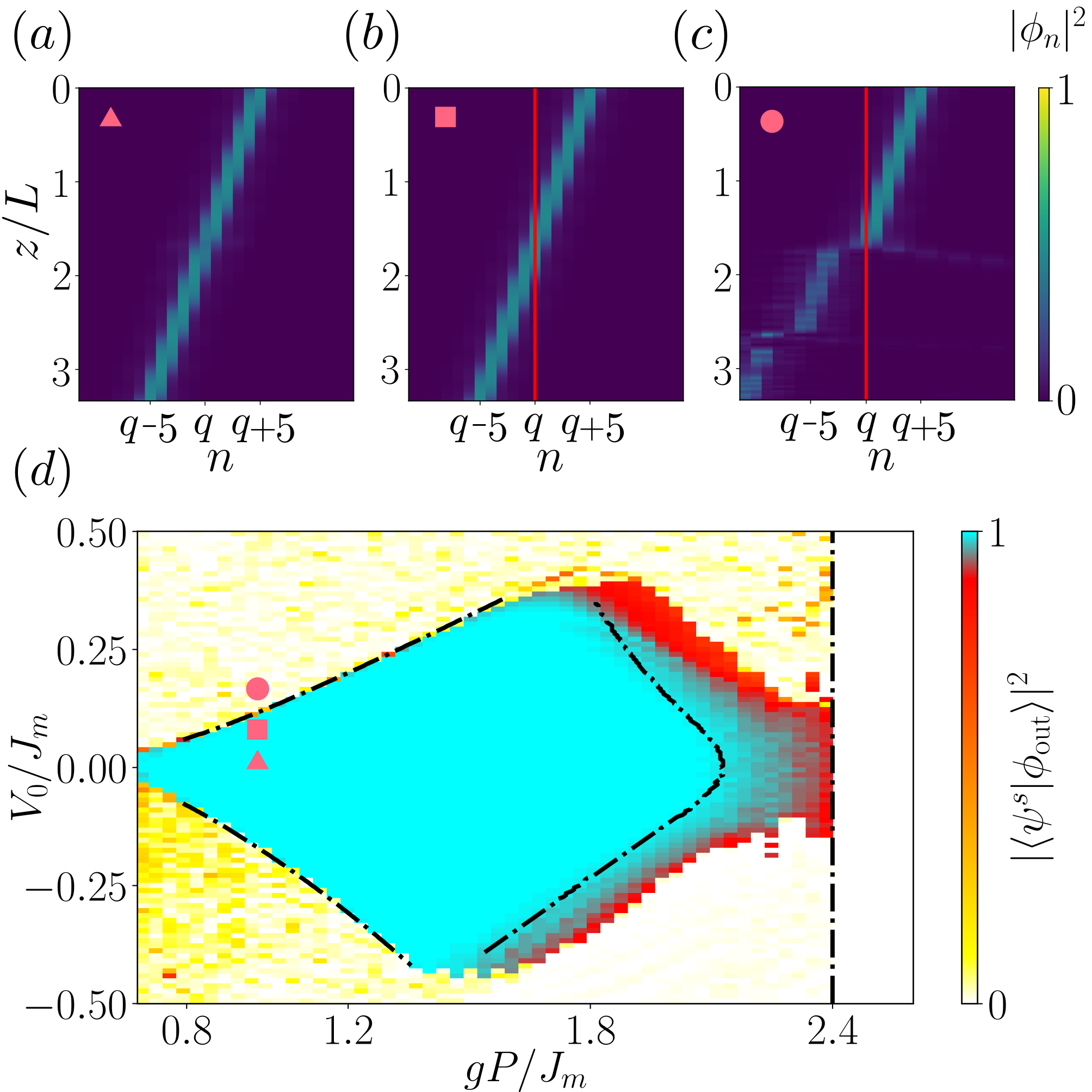}
\caption{\textbf{Numerical simulations of soliton pumping across a single defect}. ($a$-$c$) Soliton propagation for increasing defect strength ($V_0/J_m = \{0, 0.1, 0.13\}$ for ($a$) to ($c$), respectively). The red vertical line shows the defect position. Symbols mark the parameters associated with the phase diagram in ($d$). ($d$) Phase diagram showing the overlap between the numerically propagated soliton, $|\phi_{\mathrm{out}}\rangle$, at $z=10L/3$ and the instantaneous soliton, $|\psi^{s}\rangle$, centered around $q$-$5$, as a function of defect strength, $V_0/J_m$, and power, $gP/J_m$. This overlap signals the degree of perfect transmission across the defect. Black lines are nonlinear bifurcation points extracted independently from time-independent simulations. These simulations were performed for a system with $N=32$ unit cells and period $L = 6000$.}
\label{delta_theory}
\end{figure}

Here, $d$ is the average spacing between two neighboring waveguides, $\delta d$ is the modulation strength, and $L$ is the pump period. As $x_n(z) = x_{n+3}(z)$, the unit cell consists of three sites labeled $A$, $B$, and $C$ (see Fig.~\ref{schematics} ($b$)). The periodic modulation of the waveguides causes the hopping $J_n(z)$ to vary periodically in $z$ (see Fig.~\ref{schematics} ($c$)). As a consequence, the first through third bands have Chern numbers \{-1,2,-1\}, respectively~\cite{jurgensen2021quantized}.

\begin{figure*}[t]
\includegraphics[width = 17cm]{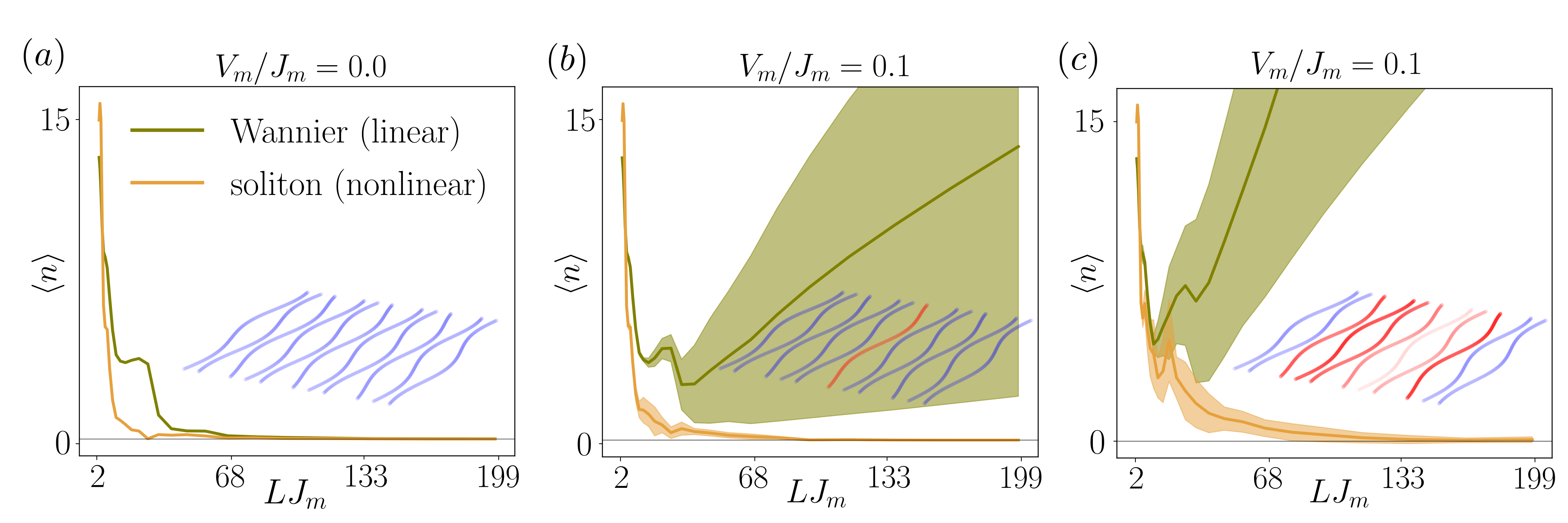}
\caption{\textbf{Comparison of topological pumping between Wannier functions and solitons}. Deviation of the center of mass from perfect pumping, $\langle n \rangle$, for Wannier functions and solitons after pumping through ($a$) a clean region ($V_m/J_m = 0$), ($b$) a disordered region with a single defect ($V_m/J_m = 0.1$) and ($c$) a disordered region with multiple defects, as a function of adiabaticity $LJ_m$. The insets show a schematic of the waveguide arrays used in the simulations. Non-zero values of $\langle n \rangle$ indicate imperfect pumping. The disordered cases are averaged over 100 disorder realizations drawn from a uniform distribution with zero mean and spread $2V_m$. The shaded region shows the $1\sigma$ region around average. The power used for soliton propagation simulation is $gP/J_m = 1.9$ and the simulations for the Wannier function and the soliton propagation were performed with $N = 350$ and $N=50$ unit cells respectively. In both cases, propagation length is $z = 10L/3$.}
\label{wannier_vs_solitons}
\end{figure*}

To investigate the soliton transport behavior, we numerically calculate the localized nonlinear eigenstates of Eq.~\ref{DNLSE} (i.e., solitons) that bifurcate from the lowest band, in the perfectly periodic case, using self-consistent iteration~\cite{jurgensen2022chern,jorg2025optical}. We then time-evolve the soliton using an adaptive Runge-Kutta method (DOP853). In the clean case ($V_n = 0$), the soliton is pumped to the left by one unit cell after one full period $L$, consistent with the Chern number of the lowest band being $-1$ (see Fig.~\ref{delta_theory} ($a$)). To investigate the effect of disorder, we first consider the special case of transport through a single defect:
\begin{equation}
\label{single_defect}
V_n = V_0\delta_{qn},
\end{equation}
where $V_0$ is the strength of the defect at site $q$. For a sufficiently small defect strength, we observe that the soliton successfully pumps through it and maintains its shape perfectly, as shown in Fig.~\ref{delta_theory} ($b$), despite slight variations in its wave function and center of mass position when near the defect, compared to the clean case. This behavior is observed until $V_0$ increases beyond a critical threshold, above which the soliton is unable to pump across it (see Fig.~\ref{delta_theory} ($c$)). 

From these propagation simulations, we extract a phase diagram showing the overlap between the pumped soliton and the instantaneous soliton at the output position, which is centered roughly five sites from the defect site ($q$-$5$), as a function of power ($gP/J_m$) and defect strength ($V_0/J_m$) (see Fig.~\ref{delta_theory} ($d$)). The phase diagram clearly exhibits a quantized transport region (shown in cyan). Surprisingly, for a fixed defect strength quantized pumping is \textit{induced} by increasing the soliton power (i.e. the photon interaction strength). The uniformity of the cyan region is a direct consequence of the quantization of transport: small perturbations cannot make incremental changes in the soliton's displacement once it has passed through the scattering region. In addition, clearly visible is the high power trapped regime $gP/J_m > 2.4$ (shown in white), in which the self-induced potential of the soliton prohibits movement away from its original site \cite{jurgensen2021quantized}. Hence, this region shows zero transport and is independent of the defect strength. 

\begin{figure*}[t]
\includegraphics[width = 17cm]{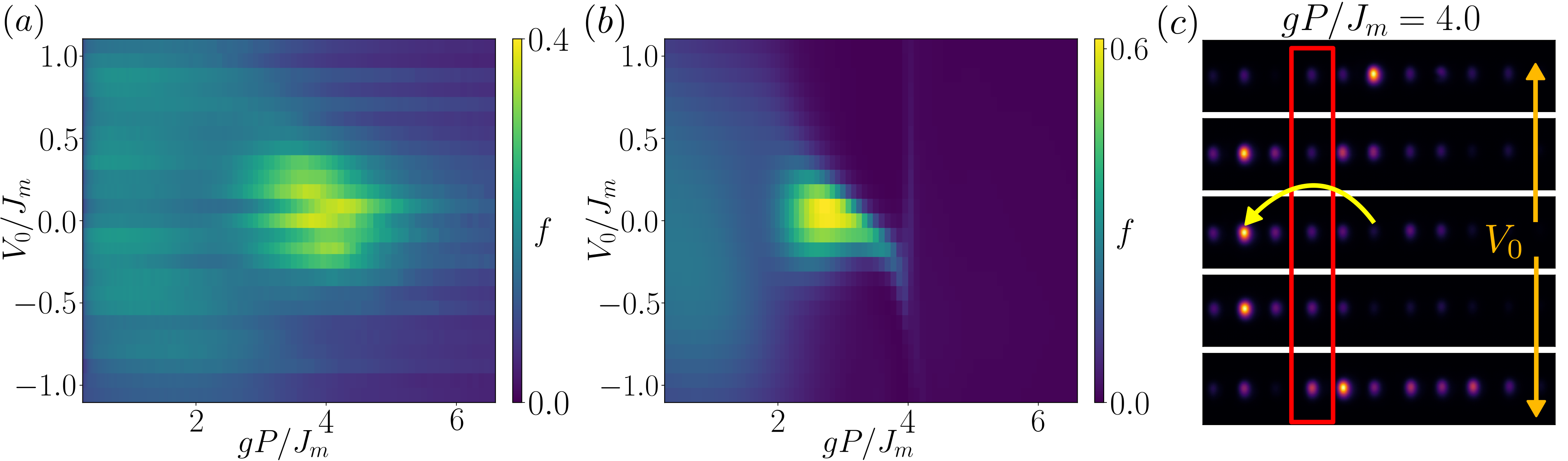}
\caption{\textbf{Experimental results for soliton pumping through a single defect.} ($a$) Experimentally obtained phase diagram showing the fraction of input power $f$ transferred to the target waveguide at the end of pumping, as a function of defect strength, $V_0/J_m$, and nonlinearity, $gP/J_m$. ($b$) Numerically calculated phase diagram. ($c$) Raw images of the output facet for $V_0/J_m = 1.06$, $V_0/J_m = 0.26$, $V_0/J_m = 0$, $V_0/J_m = -0.18$, $V_0/J_m = -0.97$ from top to bottom. The red rectangle highlights the defect site; the yellow arrow indicates the waveguides from which and to which the soliton is pumped.}
\label{delta_exp}
\end{figure*}

The breakdown of quantization (i.e., the phase boundaries in Fig.~\ref{delta_theory} ($d$)) has been shown previously to arise from nonlinear bifurcations appearing in the soliton's path during pumping~\cite{jurgensen2021quantized,jurgensen2023quantized}. We demonstrate that this is also the case in the presence of a defect by using the Newton-Raphson algorithm to find the solitons near the defect and then linearize around these solutions as part of the linear stability analysis (see supplemental material sect.~\ref{linear_stability_analysis} and ~\cite{jurgensen2021quantized,jorg2025optical,kevrekidis2009discrete}). We observe that for low powers $gP/J_m\sim 0.8$-$1.5$, the stable solitons are annihilated in a nonlinear bifurcation (see supplemental material Fig.~\ref{low_power_bifurcations} and Ref.~\cite{palmero2008solitons}), heralded by the closing of the gap in the linear stability spectrum at the phase boundary shown as black dot-dashed lines in Fig.~\ref{delta_theory} ($d$). At higher powers ($gP/J_m \sim 1.5$ - $2.1$) a separate nonlinear pitchfork bifurcation splits the previously contiguous soliton trajectory~\cite{jurgensen2021quantized,tuloup2023breakdown}, making it non-contiguous. The soliton is able to ``jump" across the stable arms of the pitchfork bifurcation with radiative losses depending on the severity of the path splitting (see the supplemental material Fig.~\ref{high_power_bifurcations}) rendering the soliton pumping non-quantized. This is further detailed in supplemental material sect.~\ref{low_power_bifurcations_and_phase_boundary} and~\ref{high_power_bifurcations_and_phase_boundary}.

We now include the more general case of a scattering region in our discussion, instead of just a single defect:
\begin{equation}
\label{disorder_pot}
V_n = \sum^{q+l}_{p = q-l}V_p\delta_{pn},
\end{equation}
where $V_{p}$ is uniformly distributed within the range $[-V_m,V_m]$; $q$ is the center of the disordered region; and it covers $2l+1$ sites. We start with a comparison of the pumping of Wannier functions and solitons for ($a$) a clean system, ($b$) the single-defect case, and ($c$) a disordered region, shown in Fig.~\ref{wannier_vs_solitons}, respectively. For all three cases, we compute the center of mass as a function of increasing adiabaticity $L J_m$. 

The clean case ($V_m = 0$), shown in  Fig.~\ref{wannier_vs_solitons} ($a$), establishes the required level of adiabaticity for topological pumping. Soliton pumping (i.e., the nonlinear case) is significantly more robust to non-adiabatic effects introduced by finite pump period than Wannier function pumping (i.e., the linear case), as observed from the deviation of the center of mass from the target position. The difference between the two cases is a direct consequence of the large spectral gap (tunable by changing the soliton power) between the soliton and the linear bands~\cite{PhysRevLett.90.170404,pu2007adiabatic}, which increases with soliton power. By contrast, in the linear case, any amount of disorder will necessarily mix the populations of different Bloch states in the band, causing the quantization to break down~\cite{cerjan2020thouless}. This is shown for a single defect and the case of a more extended disorder ($7$ defect sites) in Fig.~\ref{wannier_vs_solitons} ($b$,$c$), respectively.      

\textit{Fabrication.}--- We fabricate coupled waveguide arrays using the femtosecond direct laser writing technique~\cite{davis1996writing,szameit2010discrete}. For Thouless pumping structures, waveguides have an average spacing $d = 20~\mathrm{\mu m}$. The waveguides are spatially modulated along $z$ according to Eq.~\ref{waveguide_pumping}, with $\delta d = 1.25 \mathrm{\mu m}$. As a result, the hopping strength between neighboring waveguides takes the following form:
\begin{equation}
J_n(z) = Ae^{-\alpha\left(d+\sqrt{3}\delta d\cos\left( 2\pi(n-1)/3 +  2\pi z/L \right)\right)},
\end{equation}
where $A = 6.56~(\mathrm{mm})^{-1}$ and $\alpha = 0.24~(\mu \mathrm{m})^{-1}$ are parameters determined using the experimental characterization detailed in supplemental material sect.~\ref{experimental_setup_and_waveguide_parameter_characterization}. Using these parameters, the maximum hopping strength for our experiments turns out to be  $J_m = 0.09~(\mathrm{mm})^{-1}$

We introduce a defect in the Thouless pumps (represented by Eq.~\ref{single_defect}), by increasing (decreasing) the laser writing speed for a single waveguide in the bulk, which tunes the effective propagation constant (i.e., on-site energy) of the waveguide. This increase (decrease) in writing speed manifests itself as a negative (positive) defect. In supplemental material sect.~\ref{experimental_setup_and_waveguide_parameter_characterization}, we explain the calibration of the writing speed with the defect strength. Disordered regions are fabricated by changing the writing speeds of multiple adjacent waveguides.

\begin{figure*}[ht]
\includegraphics[width=17cm]{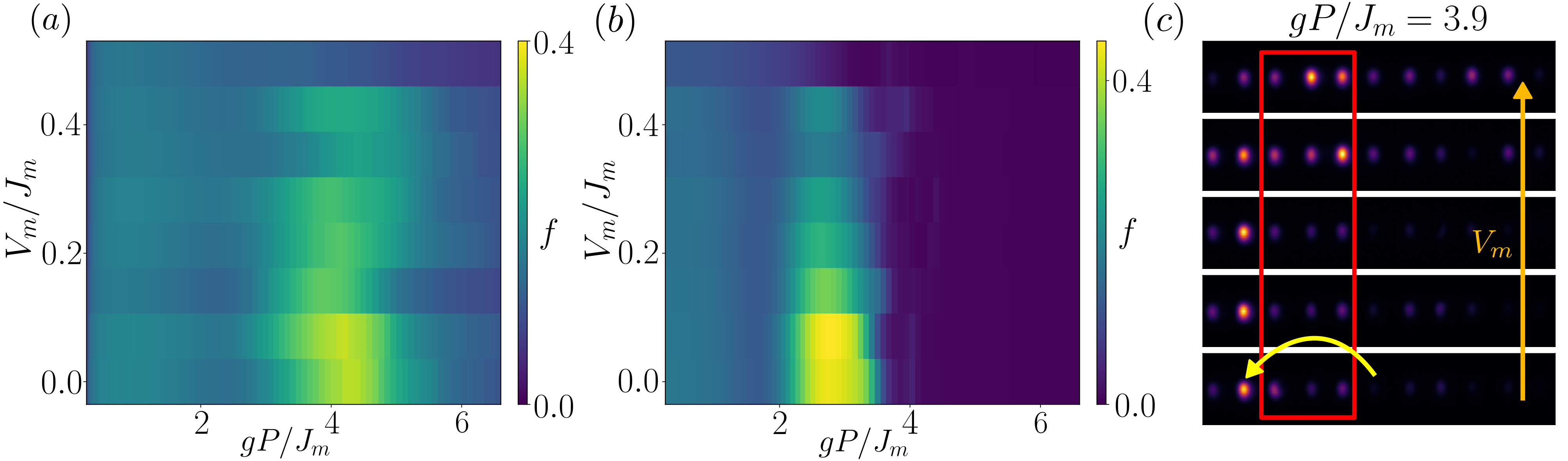}
\caption{\textbf{Experimental results for soliton pumping through a disordered region.}  ($a$,$b$) Similar to Fig.~\ref{delta_exp}, but for a three-site disordered region as opposed to a single defect. Shown is the average over six disorder realizations for each $V_m/J_m$. ($c$) Raw images of the output facet for $V_m/J_m = 0.0$, $V_m/J_m = 0.07$, $V_m/J_m = 0.14$, $V_m/J_m = 0.42$, $V_m/J_m = 0.49$, from bottom to top. The red rectangle
highlights the disordered region consisting of three sites; the yellow arrow indicates the waveguides from which and to which the soliton is pumped.}
\label{dis_exp}
\end{figure*}

\textit{Measurement.}--- To probe soliton pumping experimentally, we use an ultrafast fiber laser (Menlo Bluecut) with pulses centered on wavelength $1030~\mathrm{nm}$; the pulses are temporally stretched from $\sim260~\mathrm{fs}$ to $\sim 2~\mathrm{ps}$ and downchirped in order to mitigate spectral broadening due to self-phase modulation (see Ref. \cite{doi:10.1126/science.aba8725} for details). We inject the beam into a single waveguide, two sites away from the defect (indexed $n$=$q$+$2$), and image the output facet of the waveguide array. Provided a sufficiently high input power, a single-site excitation has a strong overlap with the soliton centered on that site. Hence, a soliton is excited and pumped along the horizontal direction as it propagates along $z$ through the waveguide array. The full length of the pump in experiments is chosen to be $z=4L/3$ ($L=57.15~\mathrm{mm}$) such that the soliton is pumped by 4 sites and centered on the site $q$-$2$ at the end of propagation, when it exits the sample. 

To gauge the impact of the defect at site $q$, we measure the fraction of output power, $f$, transferred to site $q$-$2$ as a function of the defect strength, $V_0/J_{m}$, and the laser power, $gP/J_m$ (For our samples $g = 0.11~(\mathrm{mm}~\mathrm{mW})^{-1}$, see supplemental material sect.~\ref{experimental_setup_and_waveguide_parameter_characterization} for further details). The experimental and numerical results are shown side by side in Fig.~\ref{delta_exp} ($a,b$), respectively. We observe a strong increase in $f$ for the power range $gP/J_m \sim 3$-$5$, which corresponds to the soliton successfully pumping through the defect, as seen in Fig.~\ref{delta_exp} ($c$). This is consistent with the predictions from the numerical simulations performed for the experimentally realistic case (see Fig. \ref{delta_exp} ($a$)) where we see similar behavior, though for slightly lower power thresholds due to optical losses. We explain these results as follows. At low power, the soliton is spatially broad; therefore, single-site excitation does not efficiently excite it, and hence there is no robust pumping across the defect. As power is increased, the soliton becomes more localized and is therefore excited more faithfully, and is pumped robustly across the defect. For samples with increasing defect strength, the defect-induced nonlinear bifurcation sets in, which destroys the robust pumping of the soliton through the defect. This observation therefore constitutes the realization of protected pumping across the defect. At even higher power, self-trapping occurs and the soliton remains in the site at which it was excited for the entire propagation.

Next, we study the pumping of solitons through a disordered region consisting of three sites. Their on-site energies $V_{p}$ are uniformly distributed within the range $[-V_m,V_m]$. For each value of $V_m$, we fabricate six samples with different defect configurations and measure $f$. The ensemble-averaged results obtained from the measurement of these samples are presented in Fig.~\ref{dis_exp} ($a$). Similarly to the case of a single defect, numerical simulations using experimental parameters are shown in Fig.~\ref{delta_exp} ($b$). We observe that there is a window of soliton powers $gP/J_m \sim 3$-$5$ where robust pumping is observed through the disordered region (see Fig.~\ref{dis_exp} ($c$)) before it breaks down for high disorder strength. This leads us to the conclusion that the essential physics of quantized pumping can likely be understood using the case of just a single defect and that the breakdown of transport arises due to nonlinear transitions that prevent the soliton from continuing its path in the pump.

\textit{Conclusions}---In this paper, we theoretically and experimentally demonstrate quantized soliton transport through defects and disorder in nonlinear Thouless pumps based on optical waveguide arrays. We found that transport in the nonlinear case (i.e., soliton motion) is much more robust than in the linear case (i.e., Wannier function motion) against both the presence of disorder and deviations from adiabaticity. We argue that the robustness of transport arises as a result of the quantization of the soliton displacement and that its breakdown is caused by the annihilation of the stable soliton during pumping at low power, while the pitch fork bifurcation at high power renders the soliton trajectory non-contiguous. This is an example of how nonlinear optical effects can surprisingly enhance topological properties, making the (linear) Thouless pump and the nonlinearity act together as greater than the sum of their parts. Thouless pumps are also conceptually similar to time-modulated non-reciprocal elements, in particular optical isolators \cite{lira2012electrically}, meaning that the robust transport seen here may aid in the conceptual design of future non-reciprocal devices under nonlinear operation.  

\textit{Acknowledgements}---We acknowledge the support of the Office of Naval Research under grant number N00014-23-1-2102, as well as the Air Force Office of Scientific Research under the MURI program, grant number FA9550-22-1-0339. The authors thank Nicholas Smith of Corning for providing Eagle XG glass samples for waveguide array fabrication. 

\newpage

\title{Supplemental Material: Quantized Pumping in disordered nonlinear Thouless pumps}

\maketitle

\newcommand{\beginsupplement}{%
    \clearpage
    \renewcommand{\thefigure}{S\arabic{figure}}
    \renewcommand{\theequation}{S\arabic{equation}}
    \renewcommand{\thetable}{S\arabic{table}}
    \renewcommand{\thesection}{S\arabic{section}}
}
\setcounter{figure}{0}

\beginsupplement

\section{Linear stability analysis}
\label{linear_stability_analysis}
To determine the spectral stability of a soliton ($u_n$) with eigenvalue ($\Lambda$), we follow Ref.~\cite{kevrekidis2009discrete} and add the following perturbation to $u_n$: 
\begin{equation}\label{perturbation}
\begin{aligned}
\phi_n(z) = e^{-i\Lambda z}(u_n &+ \epsilon (v_n + iw_n)e^{i\omega z}\\ 
&+ \epsilon (v^{*}_n + iw^{*}_n)e^{-i\omega z}),
\end{aligned}
\end{equation}
and look at the evolution of the perturbation. The equation of motion for the perturbation is obtained by inserting Eq.~\ref{perturbation} into the discrete nonlinear Sch\"odinger equation
\begin{equation}\label{DNLSE_2}
i \partial_z \phi_n(z) = \sum_{m}H_{nm}\phi_m(z)-g|\phi_n(z)|^2 \phi_n(z).
\end{equation}
Retaining the $\mathcal{O}(\epsilon)$ order terms gives the following eigenvalue equations for the perturbations:
\begin{align} 
\omega^{2} w_n &= \sum_{m}(L^{+}L^{-})_{nm} w_m,\\
\omega^{2} v_n &= \sum_{m}(L^{-}L^{+})_{nm} v_m,
\end{align}
where $L^{-} = (H - g\mathrm{Diag}((u_n)^2) - \Lambda \mathbb{I})$ and $L^{+} = (H - 3g\mathrm{Diag}((u_n)^2) - \Lambda \mathbb{I})$. $u_n$ is said to be spectrally unstable if $\omega^2<0$ for some eigenvector in the spectrum of $L^{+}L^{-}$. Given that $u_n$ is spectrally stable, the linearized spectrum of the soliton consists of eigenvalues $\omega^{2} > 0$ and $\omega^2=0$, which correspond to linear bands and defect states (including $u_n$), respectively~\cite{kevrekidis2009discrete}.

\section{Low power bifurcations and phase boundary}
\label{low_power_bifurcations_and_phase_boundary}

\begin{figure}[ht]
    \includegraphics[width=8.5cm]{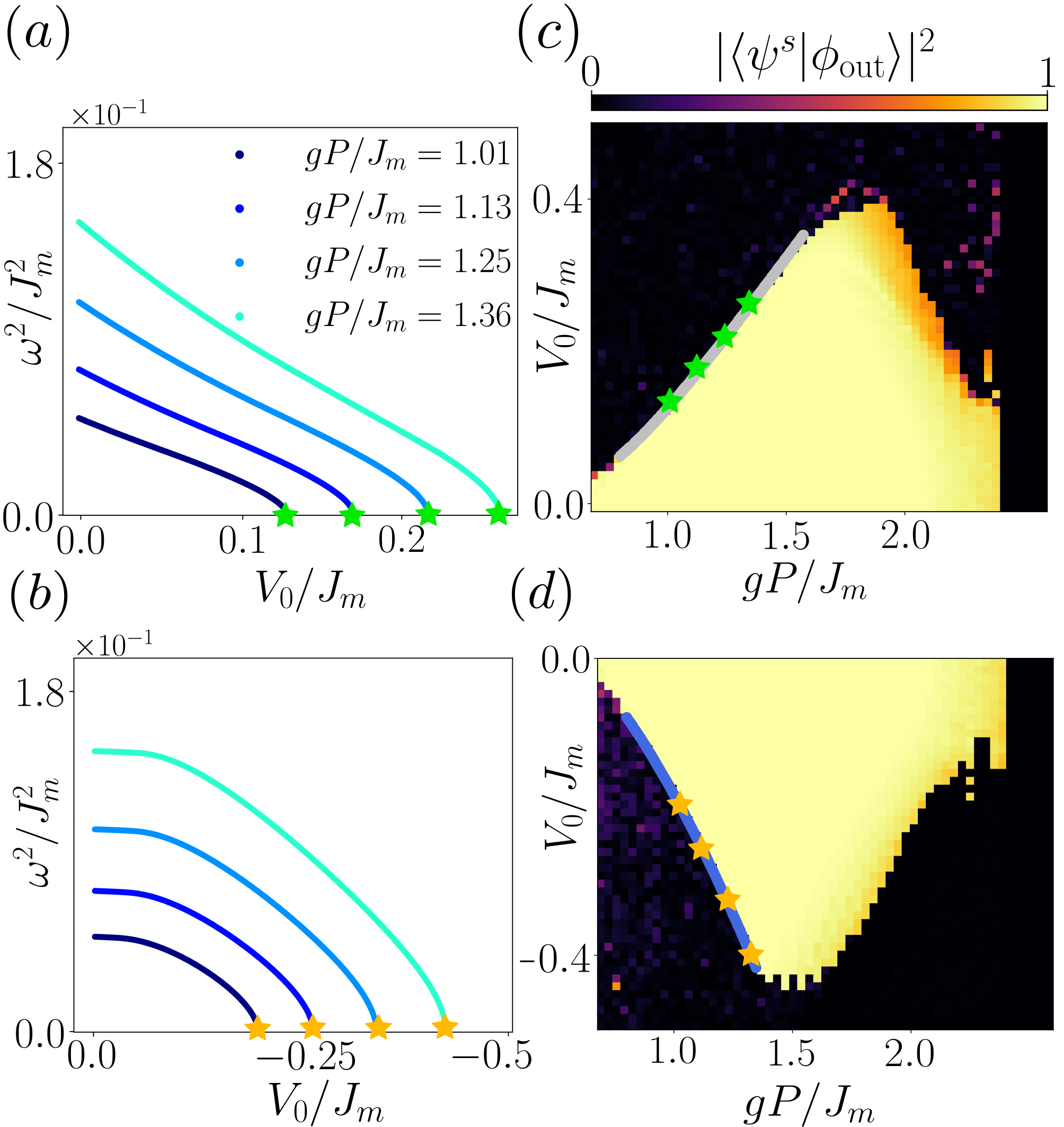}
    \caption{\textbf{Low-power phase boundaries:} Gap in the linearized spectrum of the instantaneous soliton ($\omega^2/J_m^2$) ($a$) when it is centered on site $q$ for $V_0>0$ and ($b$) when it is centered on site $q$+$3$ for $V_0<0$. Phase diagram as shown in Fig.2 ($a$) of the main text, overlaid with the calculated phase boundary along which the gap closes for ($c$) $V_0>0$ (gray line) and ($d$) $V_0<0$ (blue line). The stars show the same values as in ($a,b$).}
    \label{linear_stability}
\end{figure}

The propagation simulations shown in the main text start with a localized soliton as the initial state. The adiabatic modulation of distance between neighboring waveguides during pumping enables the propagating soliton to follow the spectrally stable instantaneous soliton given by the solution of the following equation.
\begin{align}\label{instantaneous_ham}
\Lambda^{\mathrm{inst}} u^{\mathrm{inst}}_n &= J_n(z) u^{\mathrm{inst}}_{n+1}(z) + J_{n-1}(z) u^{\mathrm{inst}}_{n-1}(z) \nonumber \\
& \quad- g(u^{\mathrm{inst}}_n)^3.
\end{align}

\begin{figure*}[ht]
    \includegraphics[width=17.9cm]{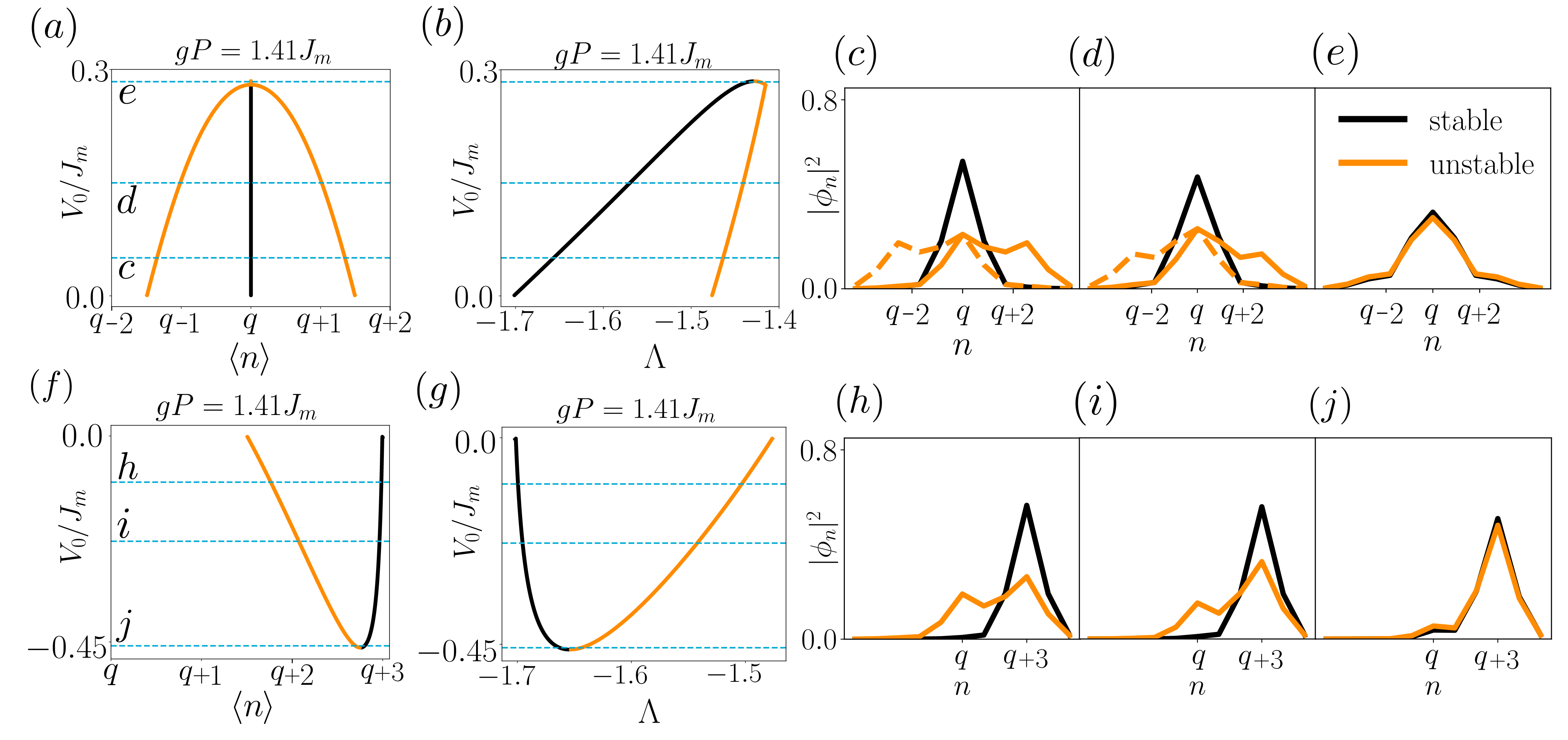}
    \caption{\textbf{Bifurcation diagrams and wavefunction shapes:} ($a$) Center of mass ($\langle n \rangle$) of instantaneous stable soliton on the defect site $q$ (black) and two unstable solitons (orange). ($b$) Corresponding eigenvalues ($\Lambda$) of the instantaneous solitons. Note that the two unstable solitons (orange) have identical eigenvalues, due to mirror symmetry about $q$. ($c$-$e$) Wavefunction of the solitons for different values of $V_0$, corresponding to the three horizontal lines in ($a,b$). ($f$-$j$) Similar to ($a$-$e$) but for $V_0<0$. Instead of two unstable solutions, only one unstable solution takes part in the nonlinear bifurcation and the stable soliton is located at $q+3$.}
    \label{low_power_bifurcations}
\end{figure*}

Here, $\Lambda^{\mathrm{inst}}$ is the instantaneous eigenvalue corresponding to the soliton $u^{\mathrm{inst}}_n$ with $\sum_n (u_n^{\mathrm{inst}})^2  = P$. For systems with a sufficiently long pump period $L$, quantized soliton pumping can occur only when the linearized spectrum of $u^{\mathrm{inst}}_n$ has a finite gap separating the linear bands and the defect states (includes the soliton $u^{\mathrm{inst}}_n$) during the Thouless pumping cycle~\cite{jurgensen2021quantized,PhysRevLett.90.170404,pu2007adiabatic}. If the gap between the linear bands and $u^{\mathrm{inst}}_n$ closes during pumping, the soliton cannot follow $u^{\mathrm{inst}}_n$ because even a slight deviation from the perfectly adiabatic limit will couple the soliton to the extended states in the linear bands, resulting in a loss of power by radiating into the extended states. 

\begin{figure}[th]
    \centering
    \includegraphics[width=8.5cm]{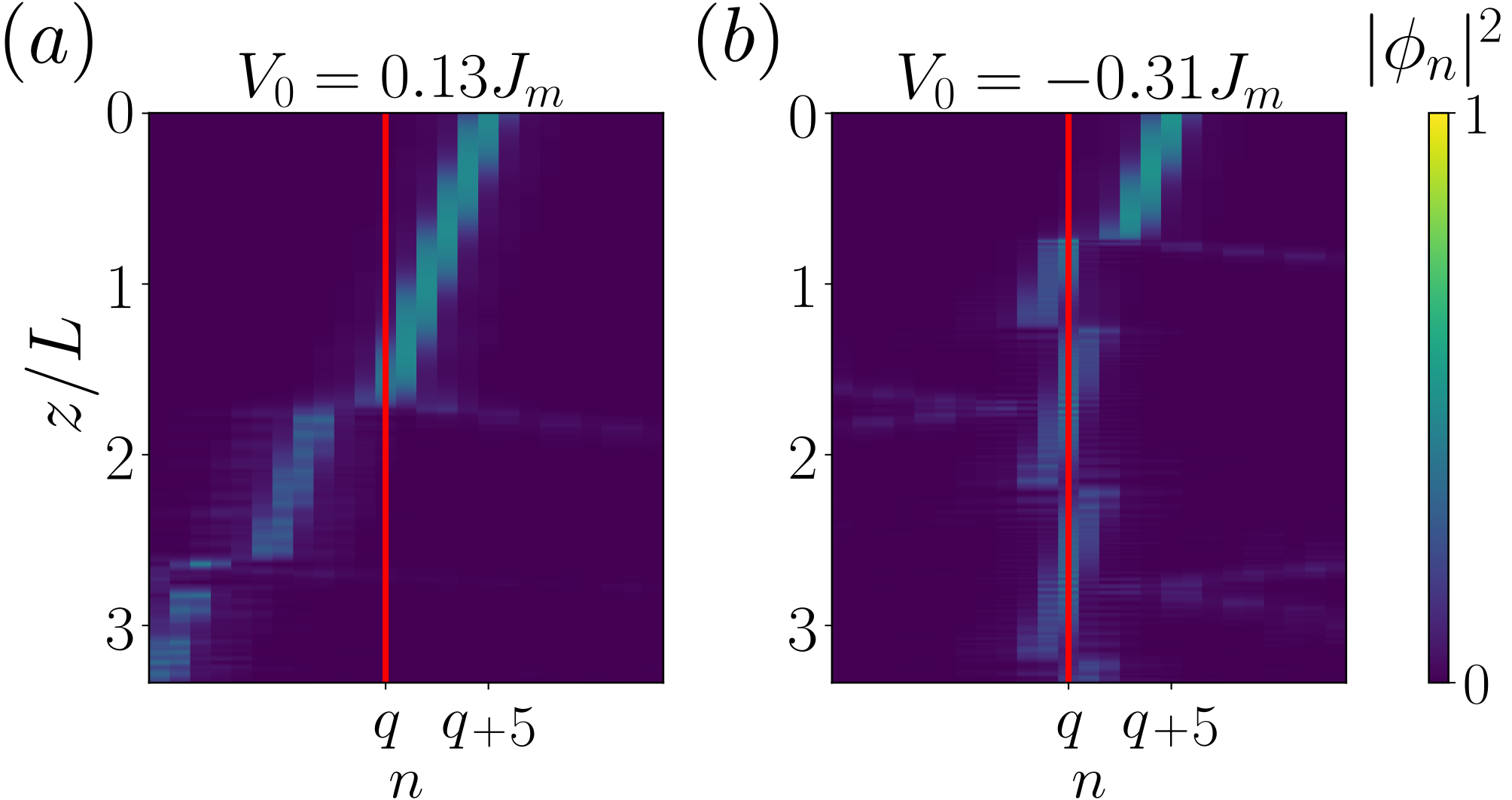}
    \caption{\textbf{Reflection of nonlinear bifurcations in propagation simulations}. ($a$) Propagation simulation for $V_0>0$, when the gap in the linearized spectrum of the instantaneous soliton at $q$, closes. ($b$) Propagation simulation for $V_0<0$, when the gap in the linearized spectrum of the instantaneous soliton at $q+3$, closes. The red vertical line marks the position of the defect site. $gP/J_m$ is 1 and 1.2 for $(a)$ and $(b)$.}
    \label{breakdown_of_pumping}
\end{figure}

We investigate the transition between phase with quantized pumping and phase without quantized pumping by analyzing the closing of the gap in the linearized spectrum of $u^{\mathrm{inst}}_n$, as a function of the input power ($P$) and the defect strength ($V_0$). It is observed that, for a fixed $gP/J_m$, there is a critical defect strength at which the gap separating the linear bands and $u^{\mathrm{inst}}_n$ closes at some point during the pumping cycle and therefore signals the breakdown of quantized pumping. Interestingly, for a positive defect ($V_0>0$) this happens when $u^{\mathrm{inst}}_n$ is exactly centered at the defect site $q$. However, for a negative defect ($V_0<0$) this gap closing occurs, when the soliton is centered around one unit cell (three sites) away from the defect site, at $q+3$. 

\begin{figure*}[ht]
    \includegraphics[width=17.9cm]{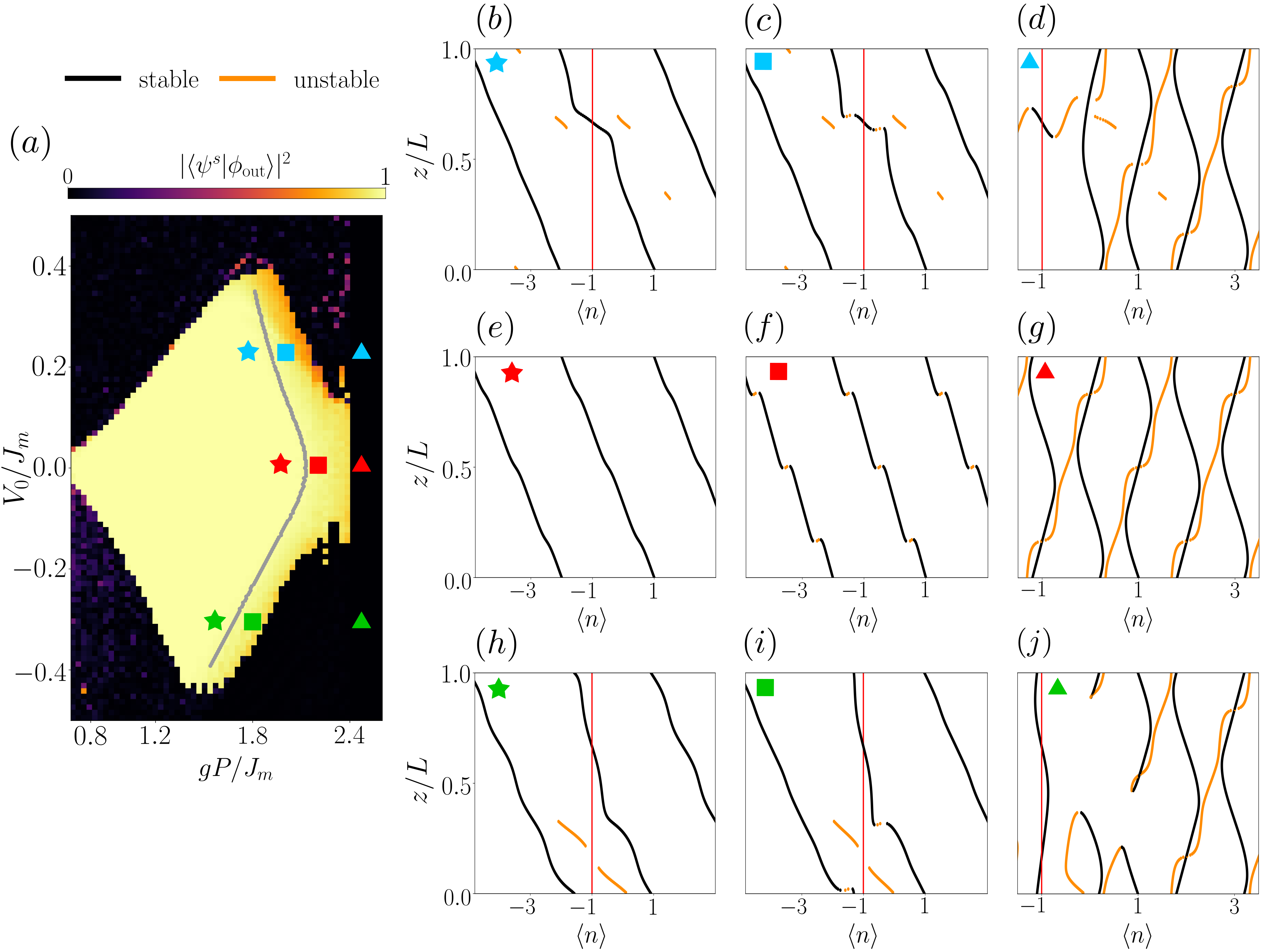}
    \caption{\textbf{High power phase boundaries:} ($a$) Phase diagram overlaid with the high power phase boundary along which the pitch fork bifurcation occurs. In ($b$), ($c$) and ($d$) we show the slices of trajectories of center of mass ($\langle n \rangle$) of instantaneous soliton solutions for $V_0>0$. In ($e$), ($f$) and ($g$) we show the slices of trajectories of center of mass ($\langle n \rangle$) of instantaneous soliton solutions for $V_0 = 0$. In ($h$), ($i$) and ($j$) we show the slices of trajectories of center of mass ($\langle n \rangle$) of instantaneous soliton solutions for $V_0 < 0$.}
    \label{high_power_bifurcations}
\end{figure*}

In Fig.~\ref{linear_stability}($a,b$), we plot the gap ($\omega^2/J^2_m$) in the linearized spectrum of $u^{\mathrm{inst}}_n$ for four different strengths of $gP/J_m$ as a function of increasing defect strength, $|V_0|$. For both positive ($V_0>0$) and negative ($V_0<0$) defects, the gap closes at a critical defect strength, $V_{\mathrm{cr}}$, marked by green and orange stars. For the defect strength above $|V_{\mathrm{cr}}|$, no quantized pumping is observed; therefore, $|V_{\mathrm{cr}}|$ determines the sharp phase boundary that separates the region in which the soliton perfectly pumps through the defect from the region in which it cannot (see Fig.~\ref{linear_stability}($c,d$)).

\begin{figure}[ht]
    \includegraphics[width=8.5cm]{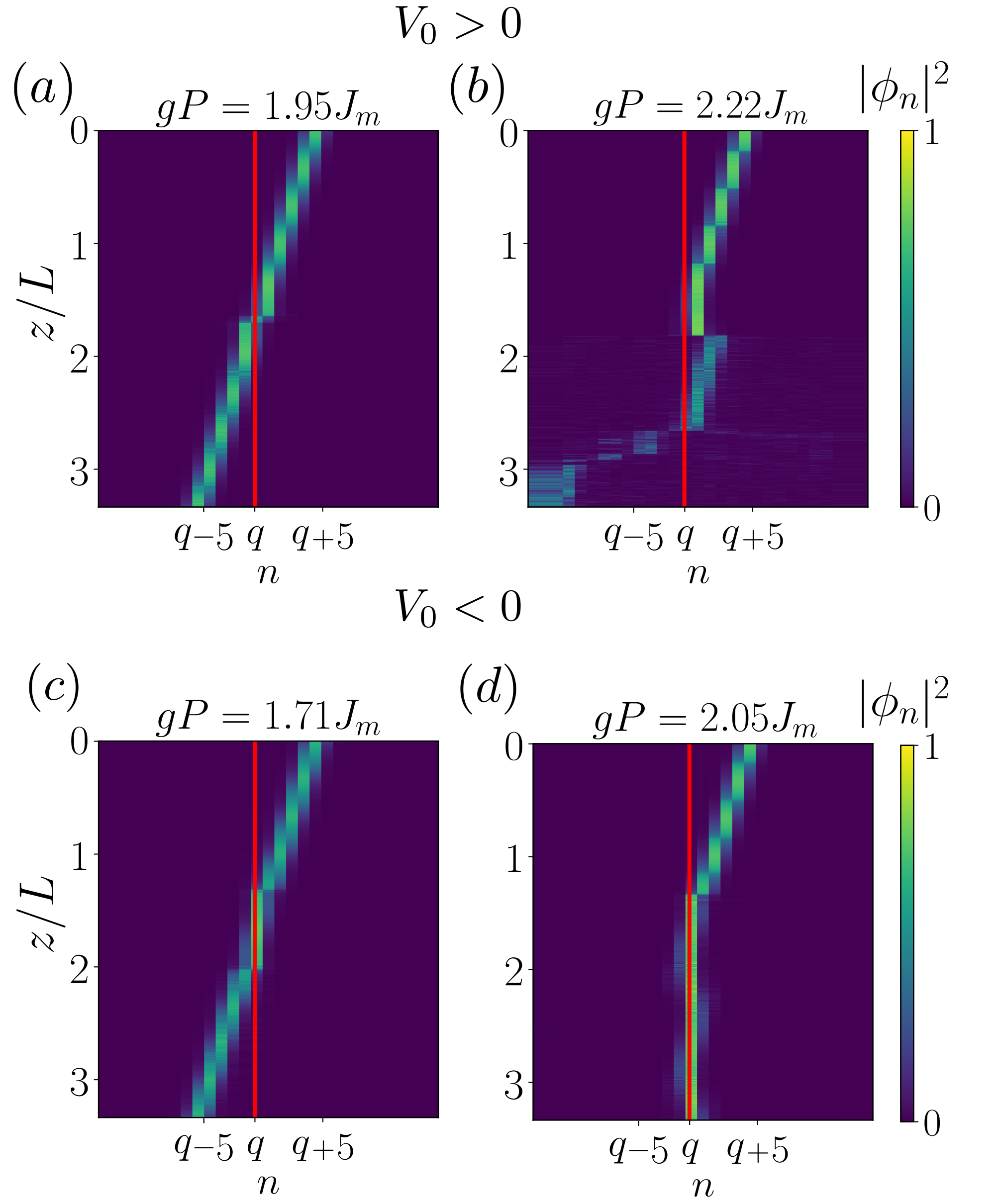}
    \caption{Propagation of an soliton across the defect in the non quantized phase for a positive defect ($a,b$) and a negative defect ($c,d$).}
    \label{pumping_breakdown_high_power}
\end{figure}

The gap closing in the linearized spectrum of $u^{\mathrm{inst}}_n$ signals the onset of defect induced nonlinear bifurcations that result in the annihilation of stable $u^{\mathrm{inst}}_n$ in the vicinity of the defect, as seen in an integer waveguide lattice with defect~\cite{palmero2008solitons}. Nonlinear bifurcations often involve more than one instantaneous soliton solution (stable or unstable) merging together as some parameter (like power, defect strength, etc.) is changed, and hence often some parameter like their respective eigenvalues or the center of masses are equal at the bifurcation point. To gain further understanding of the nature of the bifurcations that determine the phase boundaries, we present bifurcation diagrams using center of mass positions, eigenvalues, and also show the spatial shape of the involved solitons for different defect strengths in Fig.~\ref{low_power_bifurcations}.

\begin{figure}[ht]
    \includegraphics[width=8.5cm]{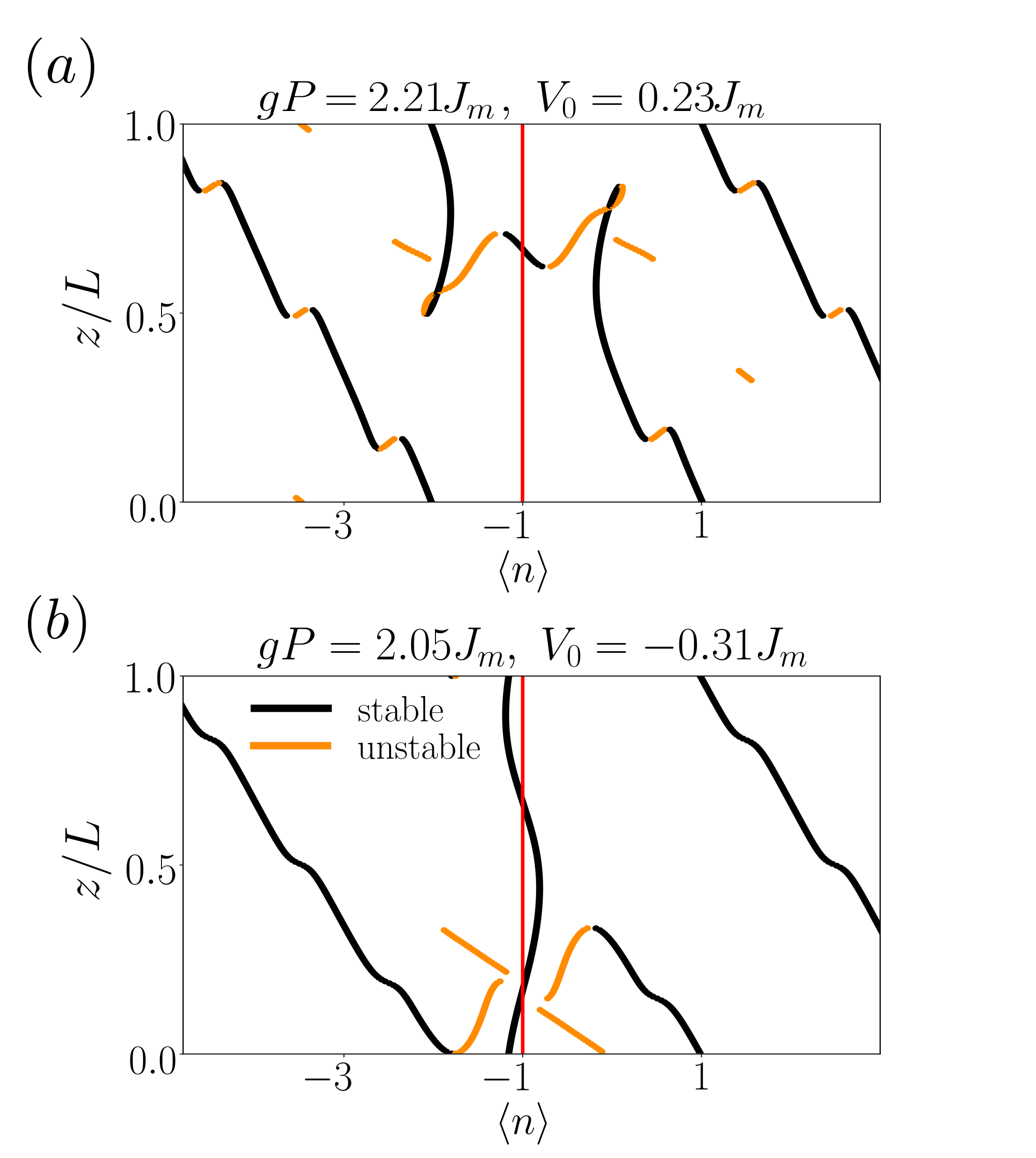}
    \caption{Instantaneous soliton trajectories in the non-quantized phase for $V_0 > 0$ and $V_0<0$ for ($a$) and ($b$), respectively. Note that no contiguous path across the defect exists.} 
    \label{high_power_bifurcations_nq_w_defect}
\end{figure}

For $V_0>0$ (upper row in Fig.~\ref{low_power_bifurcations}), we focus on three instantaneous solitons: A stable instantaneous soliton at $q$ (black), which is part of the Thouless pumping cycle, and two unstable instantaneous solutions (orange), which are not part of the Thouless pumping cycle, and whose centers of mass are left and right of the defect (see Fig.~\ref{low_power_bifurcations}($a$) and Fig.~\ref{low_power_bifurcations}($c$)-($e$) for the wavefunctions). With increasing defect strength, the two unstable instantaneous solitons start to morph into the stable instantaneous solution, eventually merging with the stable instantaneous soliton at $V_0 = V_{\mathrm{cr}}$ and resulting in its annihilation in Fig.~\ref{breakdown_of_pumping}($e$). 

When $V_0<0$ (lower row in Fig.~\ref{low_power_bifurcations}), the nature of the defect-induced nonlinear bifurcation changes. The stable instantaneous soliton (part of the Thouless pumping cycle) centered on $q+3$ merges with the unstable instantaneous soliton centered between $q+1$ and $q+2$ in a saddle-node type nonlinear bifurcation at $V_0 = V_{\mathrm{cr}}$ and hence results in its annihilation. This change in the nature of the defect-induced nonlinear bifurcation with the sign of $V_0$, can be seen in the propagation simulations shown in Fig.~\ref{breakdown_of_pumping} ($a$,$b$), where the soliton pumping breaks down when the soliton is at $q$ ($q+3$) for $V_0>0$ ($V_0<0$).

\section{High power bifurcations and phase boundary}
\label{high_power_bifurcations_and_phase_boundary}

We now turn to the high optical power phase boundary (shown by the gray line in Fig.~\ref{high_power_bifurcations} ($a$)) and first discuss the clean system ($V_0 = 0$). In this case, the occurrence of a pitch fork bifurcation at the mirror-symmetric point between lattice sites~\cite{jurgensen2021quantized} splits the contiguous trajectory of stable instantaneous solitons (see Fig.~\ref{high_power_bifurcations} ($f$)). The introduction of a defect ($V_0 \neq 0 $) breaks the mirror symmetry, which causes the bifurcation to happen at a different $z$-position as can be seen in Figs.~\ref{high_power_bifurcations} ($c$) and ($i$), and also shifts the power threshold for the bifurcation to lower powers. Although the effect of this bifurcation is clearly visible in the phase diagram in the main text that uses a nonlinear color scale, it is more subtle to spot in Fig.~\ref{high_power_bifurcations} ($a$) that uses a linear color scale. As can be seen, for power only slightly above the bifurcation threshold, the overlap between $u^{\mathrm{inst}}_n$ centered at the target site and the soliton (see Fig.~\ref{high_power_bifurcations} ($a$), Fig.~\ref{pumping_breakdown_high_power} ($a$) and ($c$)) remains high but not one, indicating non-quantized transport. We explain this as follows: For power only slightly above the bifurcation threshold, the splitting is small, and hence the propagating soliton ``jumps" from one stable branch of the pitch fork bifurcation to the other via a non-adiabatic transition while losing some power in extended states of the linear bands (also see Ref.~\cite{tuloup2023breakdown}).

At even higher optical power ($gP/J_m > 2.4$), both in the presence and absence of the defect, new stable instantaneous soliton trajectories form (see Fig.~\ref{high_power_bifurcations} ($d$),($g$) and ($j$)), which oscillate around the site in which the soliton is injected, leading to a trapped soliton with zero displacement per period. 

Note that the non-quantized phase for soliton pumping through a defect shows some distinct features different from the clean case. For $V_0>0$, we observe that, as we move further away from the bifurcation power threshold, the pitch fork bifurcation becomes more prominent and confines the soliton to the site preceding the defect site $q+1$ (see Fig.~\ref{high_power_bifurcations_nq_w_defect} ($a$)), before some power is transferred against the pumping direction to site $q+2$ as seen in Fig.~\ref{pumping_breakdown_high_power} ($b$). 

In the case of $V_0<0$, less optical power is needed to create a trapped soliton path for a soliton localized on the defect (see Fig.~\ref{high_power_bifurcations_nq_w_defect} ($b$)). Hence, during pumping, the soliton can get trapped on the defect site (see Fig.~\ref{pumping_breakdown_high_power} ($d$)), leading to sharp decrease of the propagated wavefunction with the target soliton.

Both of these phenomena result in the overlap between the soliton and the soliton at the target soliton to abruptly drop to very low values even before trapping sets in throughout the lattice. 

\begin{figure*}[ht]
    \includegraphics[width=17.9cm]{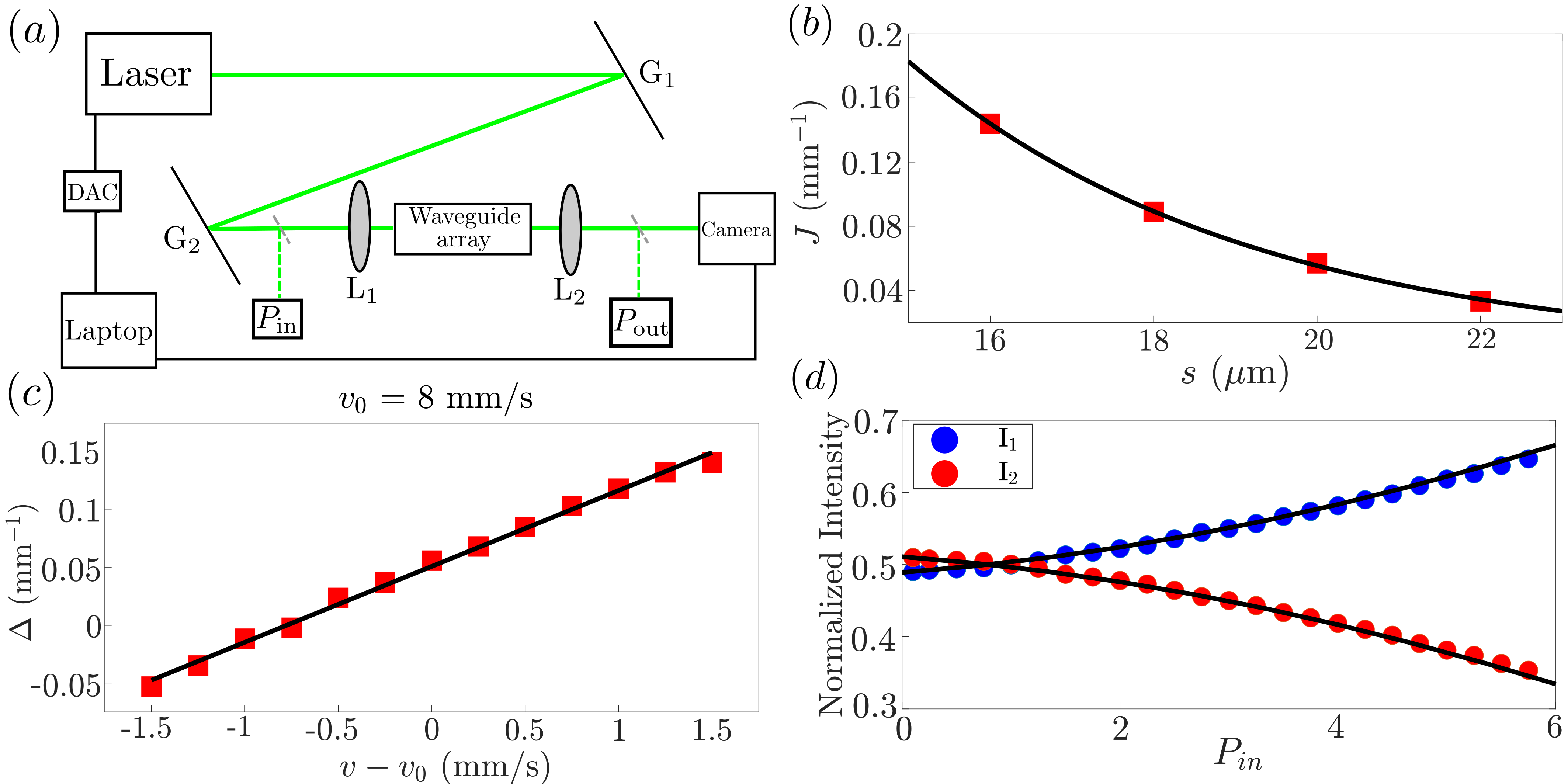}
    \caption{\textbf{Experimental setup and waveguide characterization:} ($a$) Schematic of the experimental setup used for characterization of the waveguide array parameters and the experiments presented in the main part. We use a pair of gratings $G_1$ and $G_2$ to temporally stretch the emitted laser pulses from $260~\mathrm{fs}$ to $2~\mathrm{ps}$ to avoid spectral broadening due to self-phase modulation. ($b$) Experimentally measured hopping strength as a function of waveguide separations, $s$. Each value is obtained from fitting a set of 41 couplers with varying coupling length. The solid black line is the exponential curve fitted to the measured data. ($c$) Extracted on-site potential difference as a function of change in writing speed of the waveguide. Each value is extracted from a set of 21 couplers with varying coupling length. The red squares show the experimental data and the solid black line shows the fitted line to the measured curve. ($d$) Measured values of normalized intensities in the primary ($I_{1}$) and secondary ($I_2$) waveguides of the directional coupler for increasing input power. The black curves are the obtained by solving the DNLSE for the best fit value of strength of Kerr nonlinearity ($g$).}
    \label{setup}
\end{figure*}

\section{Experimental Setup and Waveguide Parameter Characterization}
\label{experimental_setup_and_waveguide_parameter_characterization}

We use the same experimental setup for the measurements presented in the main text as well as the characterization of linear and nonlinear parameters of the waveguide arrays. A schematic of the setup is presented in Fig.~\ref{setup}$(a)$. The setup consists of a pulsed high-power fiber laser system (Menlo Bluecut) with a central wavelength of $1030~\mathrm{nm}$ and a variable pulse repetition rate (here: $5~\mathrm{kHz}$). The pulses are temporally stretched to $2~\mathrm{ps}$ to minimize spectral broadening through self-phase modulation (see Ref.~\cite{doi:10.1126/science.aba8725}). For automated power sweeps at a fixed waveguide location, we computer-control the laser power as well as the CMOS camera (Thorlabs DCC1545M-GL) imaging the output facet.

We exclusively use directional couplers to characterize the linear and nonlinear parameters. Couplers consist of a primary waveguide that extends the entire length of the glass sample and a shorter secondary waveguide (length $l$), separated by a distance $s$. When a low-power ($\sim 0.1~\mathrm{mW}$) laser beam is focused in the primary waveguide, a fraction of the power is transferred to the secondary waveguide, as a periodic function of the coupling length given by the following formula: 
\begin{equation}\label{eqFitC}
\frac{I_{2}}{I_{1} + I_{2}} = \frac{4J^2}{4J^2+\Delta^2}\sin\left(\frac{l}{2}\sqrt{4J^2+\Delta^2}\right),
\end{equation}
where the frequency and amplitude of the oscillation is dictated by the coupling strength ($J$) and the on-site potential difference ($\Delta$). To extract these parameters, we image the intensities in both waveguides at the output facet and fit it with Eq. \ref{eqFitC}. 

For each waveguide separation $s \in \{16,18,20,22\} \mu \mathrm{m}$, we use a set of 41 directional couplers with secondary waveguide lengths varying between $2$ and $62~\mathrm{mm}$, and calibrate $J$ as a function of $s$ (see Fig.~\ref{setup} ($b$)). As the overlap between the wavefunctions of neighboring waveguides decays exponentially with $s$, we expect $J(s)$ to decay exponentially with $s$ and hence fit the data with $J(s) = Ae^{-\alpha s}$ and find $A = 6.56~(mm)^{-1}$ and $\alpha = 0.24~(\mu m)^{-1}$ as the best fit values. This functional form is used for all numerical calculations presented in this manuscript. 

We extract the on-site potential difference $\Delta$ that arises between two waveguides written with different writing speeds, by fabricating sets of 21 directional couplers at fixed separation ($s = 20~\mu \mathrm{m}$) and varying coupling length ($l$). We find that the change in $\Delta$ is approximately linear in the difference in writing speed, as shown in Fig.~\ref{setup} ($c$). Notice that there is a finite on-site potential difference between ($\Delta(v=v_0)$) the primary and secondary waveguides, even in the absence of any writing speed difference. This is because the fabrication of the primary waveguide alters the neighborhood in which the secondary waveguide is written. The defect strengths mentioned in this paper are all offset by $\Delta(v=v_0)$.

For the nonlinear characterization, we use a single directional coupler ($s = 20~\mu \mathrm{m}$) and vary the power injected into the primary waveguide. We measure the power transferred to the secondary waveguide as a function of input power (see blue and red dots in Fig.~\ref{setup} ($d$)). By solving Eq.~\ref{DNLSE_2} with an initial state in which the primary waveguide is excited, we calculate the fraction of power transferred to the secondary waveguide, as a function of $P$ (shown by the black curve in Fig.~\ref{setup} ($d$)). The black curve is used to fit the experimentally obtained data for the best fit value of $g = 0.11~(\mathrm{mm}~\mathrm{mW})^{-1}$.

\bibliography{references}

@book{asboth2016short,
  title={A short course on topological insulators},
  author={Asb{\'o}th, J{\'a}nos K and Oroszl{\'a}ny, L{\'a}szl{\'o} and P{\'a}lyi, Andr{\'a}s},
  volume={919},
  year={2016},
  publisher={Springer},
  url={https://link.springer.com/book/10.1007/978-3-319-25607-8}
}

@article{lira2012electrically,
  title={Electrically driven nonreciprocity induced by interband photonic transition on a silicon chip},
  author={Lira, Hugo and Yu, Zongfu and Fan, Shanhui and Lipson, Michal},
  journal={Physical review letters},
  volume={109},
  number={3},
  pages={033901},
  year={2012},
  publisher={APS},
  url = {https://link.aps.org/doi/10.1103/PhysRevLett.109.033901}
}

@article{jurgensen2021quantized,
  title={Quantized nonlinear Thouless pumping},
  author={J{\"u}rgensen, Marius and Mukherjee, Sebabrata and Rechtsman, Mikael C},
  journal={Nature},
  volume={596},
  number={7870},
  pages={63--67},
  year={2021},
  publisher={Nature Publishing Group UK London},
  url = {https://www.nature.com/articles/s41586-021-03688-9}
}

@article{jurgensen2022chern,
  title={Chern number governs soliton motion in nonlinear thouless pumps},
  author={J{\"u}rgensen, Marius and Rechtsman, Mikael C},
  journal={Physical review letters},
  volume={128},
  number={11},
  pages={113901},
  year={2022},
  publisher={APS},
  url={https://doi.org/10.1103/PhysRevLett.128.113901}
}

@article{jurgensen2023quantized,
  title={Quantized fractional Thouless pumping of solitons},
  author={J{\"u}rgensen, Marius and Mukherjee, Sebabrata and J{\"o}rg, Christina and Rechtsman, Mikael C},
  journal={Nature Physics},
  volume={19},
  number={3},
  pages={420--426},
  year={2023},
  publisher={Nature Publishing Group UK London},
  url={https://doi.org/10.1038/s41567-022-01871-x}
}

@article{zilberberg2018photonic,
  title={Photonic topological boundary pumping as a probe of 4D quantum Hall physics},
  author={Zilberberg, Oded and Huang, Sheng and Guglielmon, Jonathan and Wang, Mohan and Chen, Kevin P and Kraus, Yaacov E and Rechtsman, Mikael C},
  journal={Nature},
  volume={553},
  number={7686},
  pages={59--62},
  year={2018},
  publisher={Nature Publishing Group UK London},
  url={https://doi.org/10.1038/nature25011}
}

@article{verbin2013observation,
  title={Observation of topological phase transitions in photonic quasicrystals},
  author={Verbin, Mor and Zilberberg, Oded and Kraus, Yaacov E and Lahini, Yoav and Silberberg, Yaron},
  journal={Physical review letters},
  volume={110},
  number={7},
  pages={076403},
  year={2013},
  publisher={APS},
  url={https://link.aps.org/doi/10.1103/PhysRevLett.110.076403}
}

@article{citro2023thouless,
  title={Thouless pumping and topology},
  author={Citro, Roberta and Aidelsburger, Monika},
  journal={Nature Reviews Physics},
  volume={5},
  number={2},
  pages={87--101},
  year={2023},
  publisher={Nature Publishing Group UK London},
  url={https://www.nature.com/articles/s42254-022-00545-0#citeas}
}

@article{cerjan2020thouless,
  title={Thouless pumping in disordered photonic systems},
  author={Cerjan, Alexander and Wang, Mohan and Huang, Sheng and Chen, Kevin P and Rechtsman, Mikael C},
  journal={Light: Science \& Applications},
  volume={9},
  number={1},
  pages={178},
  year={2020},
  publisher={Nature Publishing Group UK London},
  url={https://doi.org/10.1038/s41377-020-00408-2}
}

@article{lohse2016thouless,
  title={A Thouless quantum pump with ultracold bosonic atoms in an optical superlattice},
  author={Lohse, Michael and Schweizer, Christian and Zilberberg, Oded and Aidelsburger, Monika and Bloch, Immanuel},
  journal={Nature Physics},
  volume={12},
  number={4},
  pages={350--354},
  year={2016},
  publisher={Nature Publishing Group UK London},
  url={https://www.nature.com/articles/nphys3584}
}

@article{nakajima2016topological,
  title={Topological Thouless pumping of ultracold fermions},
  author={Nakajima, Shuta and Tomita, Takafumi and Taie, Shintaro and Ichinose, Tomohiro and Ozawa, Hideki and Wang, Lei and Troyer, Matthias and Takahashi, Yoshiro},
  journal={Nature Physics},
  volume={12},
  number={4},
  pages={296--300},
  year={2016},
  publisher={Nature Publishing Group UK London},
  url={https://www.nature.com/articles/nphys3622}
}

@article{ke2016topological,
  title={Topological phase transitions and Thouless pumping of light in photonic waveguide arrays},
  author={Ke, Yongguan and Qin, Xizhou and Mei, Feng and Zhong, Honghua and Kivshar, Yuri S and Lee, Chaohong},
  journal={Laser \& Photonics Reviews},
  volume={10},
  number={6},
  pages={995--1001},
  year={2016},
  publisher={Wiley Online Library},
  url={https://doi.org/10.1002/lpor.201600119}
}

@article{wang2022two,
  title={Two-dimensional Thouless pumping of light in photonic moir{\'e} lattices},
  author={Wang, Peng and Fu, Qidong and Peng, Ruihan and Kartashov, Yaroslav V and Torner, Lluis and Konotop, Vladimir V and Ye, Fangwei},
  journal={Nature communications},
  volume={13},
  number={1},
  pages={6738},
  year={2022},
  publisher={Nature Publishing Group UK London},
  url={https://doi.org/10.1038/s41467-022-34394-3}
}

@article{sun2022non,
  title={Non-Abelian Thouless pumping in photonic waveguides},
  author={Sun, Yi-Ke and Zhang, Xu-Lin and Yu, Feng and Tian, Zhen-Nan and Chen, Qi-Dai and Sun, Hong-Bo},
  journal={Nature Physics},
  volume={18},
  number={9},
  pages={1080--1085},
  year={2022},
  publisher={Nature Publishing Group UK London},
  url={https://doi.org/10.1038/s41567-022-01669-x}
}

@article{grinberg2020robust,
  title={Robust temporal pumping in a magneto-mechanical topological insulator},
  author={Grinberg, Inbar Hotzen and Lin, Mao and Harris, Cameron and Benalcazar, Wladimir A and Peterson, Christopher W and Hughes, Taylor L and Bahl, Gaurav},
  journal={Nature communications},
  volume={11},
  number={1},
  pages={974},
  year={2020},
  publisher={Nature Publishing Group UK London},
  url={https://doi.org/10.1038/s41467-020-14804-0}
}

@article{nakajima2021competition,
  title={Competition and interplay between topology and quasi-periodic disorder in Thouless pumping of ultracold atoms},
  author={Nakajima, Shuta and Takei, Nobuyuki and Sakuma, Keita and Kuno, Yoshihito and Marra, Pasquale and Takahashi, Yoshiro},
  journal={Nature Physics},
  volume={17},
  number={7},
  pages={844--849},
  year={2021},
  publisher={Nature Publishing Group UK London},
  url={https://doi.org/10.1038/s41567-021-01229-9}
}

@article{liu2025interplay,
  title={Interplay between disorder and topology in Thouless pumping on a superconducting quantum processor},
  author={Liu, Yu and Zhang, Yu-Ran and Shi, Yun-Hao and Liu, Tao and Lu, Congwei and Wang, Yong-Yi and Li, Hao and Li, Tian-Ming and Deng, Cheng-Lin and Zhou, Si-Yun and others},
  journal={Nature Communications},
  volume={16},
  number={1},
  pages={108},
  year={2025},
  publisher={Nature Publishing Group UK London},
  url={https://doi.org/10.1038/s41467-024-55343-2}
}

@article{PhysRevLett.45.494,
  title = {New Method for High-Accuracy Determination of the Fine-Structure Constant Based on Quantized Hall Resistance},
  author = {Klitzing, K. v. and Dorda, G. and Pepper, M.},
  journal = {Phys. Rev. Lett.},
  volume = {45},
  issue = {6},
  pages = {494--497},
  numpages = {0},
  year = {1980},
  month = {Aug},
  publisher = {American Physical Society},
  doi = {10.1103/PhysRevLett.45.494},
  url = {https://link.aps.org/doi/10.1103/PhysRevLett.45.494}
}

@article{thouless1982quantized,
  title={Quantized Hall conductance in a two-dimensional periodic potential},
  author={Thouless, David J and Kohmoto, Mahito and Nightingale, M Peter and den Nijs, Marcel},
  journal={Physical review letters},
  volume={49},
  number={6},
  pages={405},
  year={1982},
  publisher={APS},
  url = {https://link.aps.org/doi/10.1103/PhysRevLett.49.405}
}

@article{PhysRevB.23.5632,
  title = {Quantized Hall conductivity in two dimensions},
  author = {Laughlin, R. B.},
  journal = {Phys. Rev. B},
  volume = {23},
  issue = {10},
  pages = {5632--5633},
  numpages = {0},
  year = {1981},
  month = {May},
  publisher = {American Physical Society},
  doi = {10.1103/PhysRevB.23.5632},
  url = {https://link.aps.org/doi/10.1103/PhysRevB.23.5632}
}

@article{avron1983homotopy,
  title={Homotopy and quantization in condensed matter physics},
  author={Avron, Joseph E and Seiler, Ruedi and Simon, Barry},
  journal={Physical review letters},
  volume={51},
  number={1},
  pages={51},
  year={1983},
  publisher={APS},
  url = {https://link.aps.org/doi/10.1103/PhysRevLett.51.51}
}

@article{niu1985quantized,
  title={Quantized Hall conductance as a topological invariant},
  author={Niu, Qian and Thouless, Ds J and Wu, Yong-Shi},
  journal={Physical Review B},
  volume={31},
  number={6},
  pages={3372},
  year={1985},
  publisher={APS},
  url={https://doi.org/10.1103/PhysRevB.31.3372}
}

@article{fu2022nonlinear,
  title={Nonlinear Thouless pumping: solitons and transport breakdown},
  author={Fu, Qidong and Wang, Peng and Kartashov, Yaroslav V and Konotop, Vladimir V and Ye, Fangwei},
  journal={Physical Review Letters},
  volume={128},
  number={15},
  pages={154101},
  year={2022},
  publisher={APS},
  url={https://doi.org/10.1103/PhysRevLett.128.154101}
}

@article{szameit2010discrete,
  title={Discrete optics in femtosecond-laser-written photonic structures},
  author={Szameit, Alexander and Nolte, Stefan},
  journal={Journal of Physics B: Atomic, Molecular and Optical Physics},
  volume={43},
  number={16},
  pages={163001},
  year={2010},
  publisher={IOP Publishing},
  url={https://iopscience.iop.org/article/10.1088/0953-4075/43/16/163001}
}

@article{cao2024nonlinear,
  title={Nonlinear Thouless pumping of solitons across an impurity},
  author={Cao, Xuzhen and Jia, Chunyu and Hu, Ying and Liang, Zhaoxin},
  journal={Physical Review A},
  volume={110},
  number={1},
  pages={013305},
  year={2024},
  publisher={APS},
  url={https://doi.org/10.1103/PhysRevA.110.013305}
}

@article{mostaan2022quantized,
  title={Quantized topological pumping of solitons in nonlinear photonics and ultracold atomic mixtures},
  author={Mostaan, Nader and Grusdt, Fabian and Goldman, Nathan},
  journal={nature communications},
  volume={13},
  number={1},
  pages={5997},
  year={2022},
  publisher={Nature Publishing Group UK London},
  url={https://doi.org/10.1038/s41467-022-33478-4}
}

@book{kevrekidis2009discrete,
  title={The discrete nonlinear Schr{\"o}dinger equation: mathematical analysis, numerical computations and physical perspectives},
  author={Kevrekidis, Panayotis G},
  volume={232},
  year={2009},
  publisher={Springer Science \& Business Media},
  url={https://link.springer.com/book/10.1007/978-3-540-89199-4}
}

@article{davis1996writing,
  title={Writing waveguides in glass with a femtosecond laser},
  author={Davis, K Miura and Miura, Kiyotaka and Sugimoto, Naoki and Hirao, Kazuyuki},
  journal={Optics letters},
  volume={21},
  number={21},
  pages={1729--1731},
  year={1996},
  publisher={Optical Society of America},
  url={https://opg.optica.org/ol/abstract.cfm?URI=ol-21-21-1729}
}

@article{PhysRevX.14.021049,
  title = {Interactions Enable Thouless Pumping in a Nonsliding Lattice},
  author = {Viebahn, Konrad and Walter, Anne-Sophie and Bertok, Eric and Zhu, Zijie and G\"achter, Marius and Aligia, Armando A. and Heidrich-Meisner, Fabian and Esslinger, Tilman},
  journal = {Phys. Rev. X},
  volume = {14},
  issue = {2},
  pages = {021049},
  numpages = {14},
  year = {2024},
  month = {Jun},
  publisher = {American Physical Society},
  doi = {10.1103/PhysRevX.14.021049},
  url = {https://link.aps.org/doi/10.1103/PhysRevX.14.021049}
}

@article{
doi:10.1126/science.aba8725,
author = {Sebabrata Mukherjee  and Mikael C. Rechtsman },
title = {Observation of Floquet solitons in a topological bandgap},
journal = {Science},
volume = {368},
number = {6493},
pages = {856-859},
year = {2020},
doi = {10.1126/science.aba8725},
URL = {https://www.science.org/doi/abs/10.1126/science.aba8725}}

@article{PhysRevLett.109.106402,
  title = {Topological States and Adiabatic Pumping in Quasicrystals},
  author = {Kraus, Yaacov E. and Lahini, Yoav and Ringel, Zohar and Verbin, Mor and Zilberberg, Oded},
  journal = {Phys. Rev. Lett.},
  volume = {109},
  issue = {10},
  pages = {106402},
  numpages = {5},
  year = {2012},
  month = {Sep},
  publisher = {American Physical Society},
  doi = {10.1103/PhysRevLett.109.106402},
  url = {https://link.aps.org/doi/10.1103/PhysRevLett.109.106402}
}

@article{walter2023quantization,
  title={Quantization and its breakdown in a Hubbard--Thouless pump},
  author={Walter, Anne-Sophie and Zhu, Zijie and G{\"a}chter, Marius and Minguzzi, Joaqu{\'\i}n and Roschinski, Stephan and Sandholzer, Kilian and Viebahn, Konrad and Esslinger, Tilman},
  journal={Nature Physics},
  volume={19},
  number={10},
  pages={1471--1475},
  year={2023},
  publisher={Nature Publishing Group UK London},
  url={https://doi.org/10.1038/s41567-023-02145-w}
}

@article{jorg2025optical,
  title={Optical control of topological end states via soliton formation in a 1D lattice},
  author={J{\"o}rg, Christina and J{\"u}rgensen, Marius and Mukherjee, Sebabrata and Rechtsman, Mikael C},
  journal={Nanophotonics},
  volume={14},
  number={6},
  pages={769--775},
  year={2025},
  publisher={De Gruyter},
  url={https://www.degruyterbrill.com/document/doi/10.1515/nanoph-2024-0401/html}
}

@article{palmero2008solitons,
  title={Solitons in one-dimensional nonlinear Schr{\"o}dinger lattices with a local inhomogeneity},
  author={Palmero, F and Carretero-Gonz{\'a}lez, R and Cuevas, J and Kevrekidis, PG and Krolikowski, Wieslaw},
  journal={Physical Review E—Statistical, Nonlinear, and Soft Matter Physics},
  volume={77},
  number={3},
  pages={036614},
  year={2008},
  publisher={APS},
  url={https://doi.org/10.1103/PhysRevE.77.036614}
}

@article{pu2007adiabatic,
  title={Adiabatic condition for nonlinear systems},
  author={Pu, Han and Maenner, Peter and Zhang, Weiping and Ling, Hong Y},
  journal={Physical review letters},
  volume={98},
  number={5},
  pages={050406},
  year={2007},
  publisher={APS},
  url={https://doi.org/10.1103/PhysRevLett.98.050406}
}

@article{tuloup2023breakdown,
  title={Breakdown of quantization in nonlinear Thouless pumping},
  author={Tuloup, Thomas and Bomantara, Raditya Weda and Gong, Jiangbin},
  journal={New Journal of Physics},
  volume={25},
  number={8},
  pages={083048},
  year={2023},
  publisher={IOP Publishing},
  url={https://iopscience.iop.org/article/10.1088/1367-2630/acef4d}
}

@misc{jürgensen2025quantizeddynamicalpumpingdissipation,
      title={Quantized dynamical pumping via dissipation in a mechanical Thouless pump}, 
      author={Marius Jürgensen and Mikael C. Rechtsman},
      year={2025},
      url={https://arxiv.org/abs/2502.14046}, 
}

@misc{kiefer2025protectedquantumgatesusing,
      title={Protected quantum gates using qubit doublons in dynamical optical lattices}, 
      author={Yann Kiefer and Zijie Zhu and Lars Fischer and Samuel Jele and Marius Gächter and Giacomo Bisson and Konrad Viebahn and Tilman Esslinger},
      year={2025},
      url={https://arxiv.org/abs/2507.22112}, 
}

@article{4d5s-n4gn,
  title = {Multiband Fractional Thouless Pumps},
  author = {J\"urgensen, Marius and Steiner, Jacob and Refael, Gil and Rechtsman, Mikael C.},
  journal = {Phys. Rev. Lett.},
  volume = {135},
  issue = {16},
  pages = {166601},
  numpages = {6},
  year = {2025},
  month = {Oct},
  publisher = {American Physical Society},
  doi = {10.1103/4d5s-n4gn},
  url = {https://link.aps.org/doi/10.1103/4d5s-n4gn}
}

@article{96f5-qszj,
  title = {Nonlinearity-Induced Fractional Thouless Pumping of Solitons},
  author = {Tao, Yu-Liang and Zhang, Yongping and Xu, Yong},
  journal = {Phys. Rev. Lett.},
  volume = {135},
  issue = {9},
  pages = {097202},
  numpages = {6},
  year = {2025},
  month = {Aug},
  publisher = {American Physical Society},
  doi = {10.1103/96f5-qszj},
  url = {https://link.aps.org/doi/10.1103/96f5-qszj}
}

@misc{tao2024nonlinearityinducedthoulesspumpingsolitons,
      title={Nonlinearity-induced Thouless pumping of solitons}, 
      author={Yu-Liang Tao and Jiong-Hao Wang and Yong Xu},
      year={2024},
      eprint={2409.19515},
      archivePrefix={arXiv},
      primaryClass={nlin.PS},
      url={https://arxiv.org/abs/2409.19515}, 
}

@article{PhysRevB.91.064201,
  title = {Topological pumping over a photonic Fibonacci quasicrystal},
  author = {Verbin, Mor and Zilberberg, Oded and Lahini, Yoav and Kraus, Yaacov E. and Silberberg, Yaron},
  journal = {Phys. Rev. B},
  volume = {91},
  issue = {6},
  pages = {064201},
  numpages = {6},
  year = {2015},
  month = {Feb},
  publisher = {American Physical Society},
  doi = {10.1103/PhysRevB.91.064201},
  url = {https://link.aps.org/doi/10.1103/PhysRevB.91.064201}
}

@article{PhysRevLett.134.093801,
  title = {Thouless Pumping in a Driven-Dissipative Kerr Resonator Array},
  author = {Ravets, S. and Pernet, N. and Mostaan, N. and Goldman, N. and Bloch, J.},
  journal = {Phys. Rev. Lett.},
  volume = {134},
  issue = {9},
  pages = {093801},
  numpages = {6},
  year = {2025},
  month = {Mar},
  publisher = {American Physical Society},
  doi = {10.1103/PhysRevLett.134.093801},
  url = {https://link.aps.org/doi/10.1103/PhysRevLett.134.093801}
}

@article{PhysRevA.107.033501,
  title = {Observation of topological pumping of a defect state in a Fock photonic lattice},
  author = {Wu, Chaohua and Liu, Weijie and Jia, Yuechen and Chen, Gang and Chen, Feng},
  journal = {Phys. Rev. A},
  volume = {107},
  issue = {3},
  pages = {033501},
  numpages = {8},
  year = {2023},
  month = {Mar},
  publisher = {American Physical Society},
  doi = {10.1103/PhysRevA.107.033501},
  url = {https://link.aps.org/doi/10.1103/PhysRevA.107.033501}
}

@article{doi:10.1073/pnas.2411793121,
author = {Kai Yang  and Qidong Fu  and Henrique C. Prates  and Peng Wang  and Yaroslav V. Kartashov  and Vladimir V. Konotop  and Fangwei Ye },
title = {Observation of Thouless pumping of light in quasiperiodic photonic crystals},
journal = {Proceedings of the National Academy of Sciences},
volume = {121},
number = {47},
pages = {e2411793121},
year = {2024},
doi = {10.1073/pnas.2411793121},
URL = {https://www.pnas.org/doi/abs/10.1073/pnas.2411793121}}

@article{PhysRevResearch.6.023010,
  title = {Realizing efficient topological temporal pumping in electrical circuits},
  author = {Stegmaier, Alexander and Brand, Hauke and Imhof, Stefan and Fritzsche, Alexander and Helbig, Tobias and Hofmann, Tobias and Boettcher, Igor and Greiter, Martin and Lee, Ching Hua and Bahl, Gaurav and Szameit, Alexander and Kie\ss{}ling, Tobias and Thomale, Ronny and Upreti, Lavi K.},
  journal = {Phys. Rev. Res.},
  volume = {6},
  issue = {2},
  pages = {023010},
  numpages = {12},
  year = {2024},
  month = {Apr},
  publisher = {American Physical Society},
  doi = {10.1103/PhysRevResearch.6.023010},
  url = {https://link.aps.org/doi/10.1103/PhysRevResearch.6.023010}
}

@article{sun2024two,
  title={Two-dimensional non-Abelian Thouless pump},
  author={Sun, Yi-Ke and Shan, Zhong-Lei and Tian, Zhen-Nan and Chen, Qi-Dai and Zhang, Xu-Lin},
  journal={Nature Communications},
  volume={15},
  number={1},
  pages={9311},
  year={2024},
  publisher={Nature Publishing Group UK London},
  url={https://doi.org/10.1038/s41467-024-53741-0}
}

@article{huang2024topological,
  title={Topological pumping induced by spatiotemporal modulation of interaction},
  author={Huang, Boning and Ke, Yongguan and Liu, Wenjie and Lee, Chaohong},
  journal={Physica Scripta},
  volume={99},
  number={6},
  pages={065997},
  year={2024},
  publisher={IOP Publishing},
  url={https://doi.org/10.1088/1402-4896/ad491e}
}

@article{tao2025thouless,
  title={Thouless Pumping of Superposition Modes in Photonic Lattices by Employing Multiple Energy Bands},
  author={Tao, Ran and Shan, Zhong-Lei and Zhang, Xu-Lin and Chen, Qi-Dai and Tian, Zhen-Nan},
  journal={Laser \& Photonics Reviews},
  pages={2402060},
  year={2025},
  publisher={Wiley Online Library},
  url={https://doi.org/10.1002/lpor.202402060}
}

@article{PhysRevA.111.033306,
  title = {Nonlinear topological pumping of edge solitons},
  author = {You, Xinrui and Xiao, Liaoyuan and Huang, Boning and Ke, Yongguan and Lee, Chaohong},
  journal = {Phys. Rev. A},
  volume = {111},
  issue = {3},
  pages = {033306},
  numpages = {12},
  year = {2025},
  month = {Mar},
  publisher = {American Physical Society},
  doi = {10.1103/PhysRevA.111.033306},
  url = {https://link.aps.org/doi/10.1103/PhysRevA.111.033306}
}

@article{PhysRevResearch.5.013020,
  title = {Correlated topological pumping of interacting bosons assisted by Bloch oscillations},
  author = {Liu, Wenjie and Hu, Shi and Zhang, Li and Ke, Yongguan and Lee, Chaohong},
  journal = {Phys. Rev. Res.},
  volume = {5},
  issue = {1},
  pages = {013020},
  numpages = {14},
  year = {2023},
  month = {Jan},
  publisher = {American Physical Society},
  doi = {10.1103/PhysRevResearch.5.013020},
  url = {https://link.aps.org/doi/10.1103/PhysRevResearch.5.013020}
}

@article{ye2025thouless,
  title={Thouless pumping of solitons in a nonlocal medium},
  author={Ye, Fangwei and Ryazhapov, Aidar H and Kartashov, Yaroslav V and Konotop, Vladimir V},
  journal={APL Photonics},
  volume={10},
  number={7},
  year={2025},
  publisher={AIP Publishing},
  url={https://doi.org/10.1063/5.0276549}
}

@article{52yh-mlfm,
  title = {Demonstration of Returning Thouless Pump in a Berry Dipole System},
  author = {Mo, Qingyang and Liang, Shanjun and Lan, Xiangke and Zhu, Jie and Zhang, Shuang},
  journal = {Phys. Rev. Lett.},
  volume = {135},
  issue = {20},
  pages = {206603},
  numpages = {6},
  year = {2025},
  month = {Nov},
  publisher = {American Physical Society},
  doi = {10.1103/52yh-mlfm},
  url = {https://link.aps.org/doi/10.1103/52yh-mlfm}
}

@article{PhysRevLett.116.200402,
  title = {Geometrical Pumping with a Bose-Einstein Condensate},
  author = {Lu, H.-I and Schemmer, M. and Aycock, L. M. and Genkina, D. and Sugawa, S. and Spielman, I. B.},
  journal = {Phys. Rev. Lett.},
  volume = {116},
  issue = {20},
  pages = {200402},
  numpages = {5},
  year = {2016},
  month = {May},
  publisher = {American Physical Society},
  doi = {10.1103/PhysRevLett.116.200402},
  url = {https://link.aps.org/doi/10.1103/PhysRevLett.116.200402}
}

@article{PhysRevLett.129.053201,
  title = {Topological Pumping in a Floquet-Bloch Band},
  author = {Minguzzi, Joaqu\'{\i}n and Zhu, Zijie and Sandholzer, Kilian and Walter, Anne-Sophie and Viebahn, Konrad and Esslinger, Tilman},
  journal = {Phys. Rev. Lett.},
  volume = {129},
  issue = {5},
  pages = {053201},
  numpages = {6},
  year = {2022},
  month = {Jul},
  publisher = {American Physical Society},
  doi = {10.1103/PhysRevLett.129.053201},
  url = {https://link.aps.org/doi/10.1103/PhysRevLett.129.053201}
}

@article{PhysRevLett.90.170404,
  title = {Nonlinear Evolution of Quantum States in the Adiabatic Regime},
  author = {Liu, Jie and Wu, Biao and Niu, Qian},
  journal = {Phys. Rev. Lett.},
  volume = {90},
  issue = {17},
  pages = {170404},
  numpages = {4},
  year = {2003},
  month = {May},
  publisher = {American Physical Society},
  doi = {10.1103/PhysRevLett.90.170404},
  url = {https://link.aps.org/doi/10.1103/PhysRevLett.90.170404}
}

@misc{bohm2025quantumtheoryfractionaltopological,
      title={Quantum theory of fractional topological pumping of lattice solitons}, 
      author={Julius Bohm and Hugo Gerlitz and Christina Jörg and Michael Fleischhauer},
      year={2025},
      url={https://arxiv.org/abs/2506.00090}, 
}

@article{RevModPhys.82.1959,
  title = {Berry phase effects on electronic properties},
  author = {Xiao, Di and Chang, Ming-Che and Niu, Qian},
  journal = {Rev. Mod. Phys.},
  volume = {82},
  issue = {3},
  pages = {1959--2007},
  numpages = {0},
  year = {2010},
  month = {Jul},
  publisher = {American Physical Society},
  doi = {10.1103/RevModPhys.82.1959},
  url = {https://link.aps.org/doi/10.1103/RevModPhys.82.1959}
}

@article{lohse2018exploring,
  title={Exploring 4D quantum Hall physics with a 2D topological charge pump},
  author={Lohse, Michael and Schweizer, Christian and Price, Hannah M and Zilberberg, Oded and Bloch, Immanuel},
  journal={Nature},
  volume={553},
  number={7686},
  pages={55--58},
  year={2018},
  publisher={Nature Publishing Group UK London},
  url={https://www.nature.com/articles/nature25000}
}

@article{fedorova2020observation,
  title={Observation of topological transport quantization by dissipation in fast Thouless pumps},
  author={Fedorova, Zlata and Qiu, Haixin and Linden, Stefan and Kroha, Johann},
  journal={Nature communications},
  volume={11},
  number={1},
  pages={3758},
  year={2020},
  publisher={Nature Publishing Group UK London},
  url={https://doi.org/10.1038/s41467-020-17510-z}
}

@article{PhysRevLett.117.213603,
  title = {Topological Pumping of Photons in Nonlinear Resonator Arrays},
  author = {Tangpanitanon, Jirawat and Bastidas, Victor M. and Al-Assam, Sarah and Roushan, Pedram and Jaksch, Dieter and Angelakis, Dimitris G.},
  journal = {Phys. Rev. Lett.},
  volume = {117},
  issue = {21},
  pages = {213603},
  numpages = {5},
  year = {2016},
  month = {Nov},
  publisher = {American Physical Society},
  doi = {10.1103/PhysRevLett.117.213603},
  url = {https://link.aps.org/doi/10.1103/PhysRevLett.117.213603}
}

@article{PhysRevA.95.063630,
  title = {Multiparticle Wannier states and Thouless pumping of interacting bosons},
  author = {Ke, Yongguan and Qin, Xizhou and Kivshar, Yuri S. and Lee, Chaohong},
  journal = {Phys. Rev. A},
  volume = {95},
  issue = {6},
  pages = {063630},
  numpages = {11},
  year = {2017},
  month = {Jun},
  publisher = {American Physical Society},
  doi = {10.1103/PhysRevA.95.063630},
  url = {https://link.aps.org/doi/10.1103/PhysRevA.95.063630}
}

@article{PhysRevB.98.245148,
  title = {Topological charge pumping in the interacting bosonic Rice-Mele model},
  author = {Hayward, A. and Schweizer, C. and Lohse, M. and Aidelsburger, M. and Heidrich-Meisner, F.},
  journal = {Phys. Rev. B},
  volume = {98},
  issue = {24},
  pages = {245148},
  numpages = {11},
  year = {2018},
  month = {Dec},
  publisher = {American Physical Society},
  doi = {10.1103/PhysRevB.98.245148},
  url = {https://link.aps.org/doi/10.1103/PhysRevB.98.245148}
}

@article{hu2024pumping,
  title={Pumping of matter wave solitons in one-dimensional optical superlattices},
  author={Hu, Xiaoxiao and Li, Zhiqiang and Chen, Ai-Xi and Luo, Xiaobing},
  journal={New Journal of Physics},
  volume={26},
  number={12},
  pages={123006},
  year={2024},
  publisher={IOP Publishing},
  url={https://iopscience.iop.org/article/10.1088/1367-2630/ad9770}
}
\bibliographystyle{apsrev4-2.bst}

\end{document}